\documentclass[aps,twocolumn,showpacs, nofootinbib,floatfix]{revtex4}

\usepackage[centertags]{amsmath}
\usepackage{amsfonts}
\usepackage{graphicx}

\newcommand{\ve}[1]{\ensuremath{\mathbf{#1}}}
\newcommand{\av}[1]{\ensuremath{\langle#1\rangle}}
\newcommand{\dd}{\ensuremath{\mathrm{d}}}
\newcommand{\be}{\begin{equation}}
\newcommand{\ee}{\end{equation}}
\newcommand{\LL}{\langle \Lambda \rangle}
\newcommand{\bea}{\begin{eqnarray}}
\newcommand{\eea}{\end{eqnarray}} 
     
\begin{document}

  \title{Growth of correlations in gravitational N-body simulations}

  \date{\today}

  \author{Thierry \surname{Baertschiger}}
  \affiliation{D\'epartement de Physique Th\'eorique, Universit\'e de
    Gen\`eve, Quai E. Ansermet 24, CH-1211 Gen\`eve, Switzerland}
  \email{Thierry.Baertschiger@physics.unige.ch}
  \author{Francesco \surname{Sylos Labini}}
  \affiliation{Laboratoire de Physique Th\'eorique, Universit\'e Paris XI,
  B\^atiment 211, F-91405 Orsay, France}
  \email{Francesco.Sylos-Labini@th.u-psud.fr}


\begin{abstract} In the gravitational evolution of a cold infinite
particle distribution, two-body interactions can be predominant at
early times: we show that, by treating the simple case of a Poisson
particle distribution in a static universe as an ensemble of isolated
two-body systems, one may capture the origin of the first non-linear
correlated structures.  The developed power-law like behavior of the
two-point correlation function is then simply related to the
functional form of the time evolved nearest-neighbor probability
distribution, whose time dependence can be computed by using Liouville
theorem for the gravitational two-body problem. We then show that a
similar dynamical evolution is also found in a large-scale ordered
distribution, which has striking similarities to the case of a
cosmological CDM simulation which we also consider.
\end{abstract}

\maketitle

  \section{Introduction}

  Non-linear gravitational clustering can be studied by means of
  N-body simulations (NBS) which compute numerically the evolution of
  a system of particles under the action of their mutual gravity. The
  gravitational many-body problem consists in the explanation of the
  time evolution of the NBS and in the theoretical understanding of
  the formation of non-linear structures. Up to now, two different
  approaches have been generally studied: on the one hand research of
  approximative solutions of the BBGKY hierarchy~\cite{peebles} and on
  the other hand statistical thermodynamics mainly developed by
  Saslaw~\cite{saslaw}.

  A main issue in the context of cosmological NBS is to relate the
  formation of non linear structures to the specific choice of initial
  conditions used: this is done in order to constraint models with
  observations of cosmic microwave background radiation
  anisotropies, which are related to the initial conditions, and of
  galaxy structures, which give instead the final configuration of
  strongly clustered matter. Standard primordial cosmological
  theoretical density fields, like the cold dark matter (CDM) case,
  are Gaussian and made of a huge number of microscopic mass
  particles, which are usually treated theoretically as a
  self-gravitating collisionless fluid
  \cite{melott90,kuhlman96,melott}: this means that the fluid must be
  dissipation-less and that two-body scattering should be small.  The
  problem then being in which limit NBS, based on particle dynamics,
  are able to reproduce the two above conditions. In this context one
  has to consider the issue of the physical role of particle
  fluctuations in the dynamics of NBS as the total energy is conserved
  during time evolution (the only mechanism of energy dissipation is
  related to local gravitational processes).

  In fact, in the discretization of a continuous density field one
  faces two important limitations corresponding to the new length
  scales which are introduced. On the one hand a relatively small
  number of particles are used: this introduces a mass scale which is
  the mass of these particles. (In typical cosmological NBS, this mass
  is of the order of a galaxy and hence many orders of magnitude
  larger than the microscopic mass of CDM particle.)  Furthermore, it
  introduces a new characteristic length scale given by the average
  distance between nearest neighbor (NN) particles $\langle \Lambda
  \rangle$.  Clearly the discretization method used should conserve
  the continuous correlations, but this is a problematic aspects of
  standard methods \cite{bsl02,bsl03,dk03,lebo,andrea}.  On the other
  hand one must regularize the gravitational force at small scales in
  order to avoid problems related to the divergence of the numerical
  integrator and remove collisional effects due to strong scattering
  between particles.  This is usually done by using a softening length
  $\epsilon$ in the gravitational potential generally defined as 
  \begin{equation}
    \phi(r) = - \frac{1}{\sqrt{\epsilon^2+r^2}} \;. 
  \end{equation}
  This is the
  second length scale introduced to numerically simulate the
  collisionless fluid.

  The question which naturally arises is then how to choose the two
  new length scales $\langle \Lambda \rangle$ and $\epsilon$: the
  first obvious condition is that they must be both smaller than the
  intrinsic characteristic scales of the continuous field (as for
  example smaller than the typical scale corresponding to the
  turn-over scale of the CDM power spectrum).  Then one has to tune
  the ratio $\eta = \langle \Lambda \rangle / \epsilon$ appropriately
  with respect to the physical problem under study.  In fact, when
  $\eta >1$ one has a larger dynamical range than the case $\eta < 1$,
  but strong scattering between nearby particles are not smoothed and
  hence one is not effectively reproducing a dynamics where particles
  play the role of collisionless fluid elements.  It is in this sense
  that one talks about the role of discreteness in NBS: that strong
  scattering between nearby particles are produced by the
  discretization and by the choice of $\eta > 1$ and they should be
  considered artificial and spurious with respect to the dynamical
  evolution of a self-gravitating fluid. This point has been
  considered in different ways and contexts by many authors,
  \emph{e.g.}~\cite{melott90, kuhlman96, melott, bottaccio, luciano,
  moore03,bk03}: they all show that discreteness has some influence on
  the formation of the structures.

  For this reason discreteness, which anyway introduces large
  fluctuations in the density field up to scales of order $\langle
  \Lambda \rangle$, may play an important role in the early stages
  of non-linear structures formation, i.e. when the average
  distance between nearby particles becomes rapidly smaller than
  $\langle \Lambda \rangle$.  How discrete effects are then
  ``exported'' toward large scales, if they are at all, is then a deep
  and difficult problem to be understood. In other words the problem
  is that of understanding whether large non-linear structures, which
  at late times contain many particles, are produced solely by the
  collisionless dynamics of a fluid and its density fluctuations or
  whether the particle collisional processes are important also on the
  long-term.  For example \cite{bottaccio} have argued that
  discreteness effects play an important role in the self-similar
  evolution of correlated structures, while the effect of NN
  interactions has been the subject of a toy model developed by
  \cite{luciano}.

  In \cite{bjsl02,slbj03} we have already considered the effects of
  discretization in the dynamics of non linear structure formation in
  several NBS with and without space expansion. We have concluded that
  the fluctuations at the smallest scales in these NBS --- i.e.
  those associated with the discreteness of the particles --- play a
  central role in the dynamics of clustering in the non-linear
  regime. This was based in particular on the fact that the
  correlations appear to be built up from the initial clustering at
  the smallest scales and that the nature of the clustering seems to
  be independent (or at most very weakly dependent) on the initial
  conditions. The theoretical understanding of the creation of these
  correlations should therefore deal with the apparently crucial role
  of the intrinsically highly fluctuating initial density field.

  In this paper we put our previous results on a firmer physical
  basis.  We study the formation of first structures in several
  NBS. As a reference example we use a cold (zero initial velocity)
  Poisson distribution as initial conditions and we consider the case
  of a non-expanding background, i.e. a static universe. In
  this case, we show that two-body interactions are enough to explain
  the evolution of the correlation function at early times, as it has
  been already noticed in \cite{bott01}. This is done by treating
  the N-body problem as an ensemble of isolated two-body systems.
  Such an approximation is justified, in the Poisson case, by the fact
  that the probability that nearby particles are mutually NN is high enough
  ($\sim 0.6$) (becoming of order one when very close particles are
  only considered) and by the fact that the NN force is the dominating
  one ~\cite{chandra43}.  Using Liouville theorem for the
  gravitational two-body problem, we can find the early evolution of
  the NN probability distribution. As this distribution can be linked
  to the conditional density and therefore to the reduced
  two-point correlation function, we also obtain their evolution at
  early times. Comparing with the results from the simulations we find
  an excellent agreement: this shows that the first structures
  observed are a consequence of two-body interactions between
  NNs. After a time of the order of the typical time scale of two-body
  interaction, this is of course not the case anymore. However we note
  that the functional behavior of the two-point correlation function
  remains unchanged at later times, while the regime of strong
  clustering increases with time.

  We then study in the same perspective three other different
  simulations in which the force is not dominated by short-scales
  contributions since the beginning. The link between the NN
  probability distribution is found to be an efficient tool to
  study the nature of the first correlations developed and the growth
  of power-law correlations when high resolution ($\eta \gg 1$) NBS are considered.


  \section{Statistical tools}
  
  A simple tool used to study clustering of a matter distribution is
  the \emph{two-point correlation function} \cite{GJS}
  $\av{n(\ve{r}_1) n(\ve{r}_2)}$ which gives the probability density
  for finding one particle around $\ve{r}_1$ and a second one around
  $\ve{r}_2$ ($n(\ve{r})$ being the microscopic mass density
  function).  In the following we will restrict ourselves to
  distributions which have a well-defined average density $n_0$ and
  are homogeneous and isotropic. In that case, the two-point
  correlation function only depends on $r_{12}=|\ve{r}_1-\ve{r}_2|$
  and the asymptotic average density is positive.  This function is
  useful to study both continuous and discrete distributions of
  matter. In the latter case, which is the case of interest here, it
  can be useful to measure averages from a point occupied by a
  particle.  For instance, one can define the \emph{conditional
    density} 
  \begin{equation}
    \av{ n(r)}_p \equiv \frac{\av{ n(\ve{0}) n(\ve{r})}}{n_0}
  \end{equation}
  for $r>0\;$; this gives the average density at a distance $r$ from
  an occupied point~\footnote{$\av{.}_p$ means that it is a conditional
  average: the origin is an occupied point.}.  It is easy to show
  that one has the following relation
  \begin{equation}
  \av{ n(r)}_p \equiv n_0 [1+ \xi(r)] \text{ for } r>0
  \end{equation}
  where
  $\xi(r)$ is the \emph{non-diagonal part of the reduced two-point
    correlation function}~\cite{GJS}.

  In order to study small-scales properties of a discrete distribution
  one may consider the \emph{nearest neighbor probability
  distribution} $\omega(r)$. This gives the probability density of the
  distance from a particle to its NN \cite{chandra43}.  Let us briefly
  discuss its relation to the average conditional density. By
  definition, the probability that, given a particle, there is another
  particle in the infinitesimal volume element $\dd V$ at distance $r$
  is %
  \begin{equation}
    p_1(r)=\av{ n(r)}_p \,\dd V.\label{eq:p_1}
  \end{equation}
  Now we only have to note that the probability $\omega(r)\, \dd r$ for a
  given particle of having a NN at a distance between $r$ and $r+\dd
  r$ is the probability of having no NN in the sphere of radius $r$
  centered on the particle multiplied by the probability of having one
  particle in the infinitesimal spherical shell around this sphere: %
  \begin{equation}
    \label{omega1}
    \omega(r) \, \dd r 
    = \left( 1 - \int_0^r\omega(s)\, \dd s \right) \cdot \av{
    n(r)}_p 4 \pi r^2 \, \dd r
  \end{equation}
  where the second part of the right hand side is the probability
  $p_1(r)$ with $\dd V = 4\pi r^2\, \dd r$.

  \section{Evolution of a Poisson distribution}

In the case of a Poisson distribution one simply has $\av{
n(r)}_p=n_0$ \cite{GJS}.  It is then easy to solve Eq. \eqref{omega1}
for $\omega(r)$. One finds \cite{chandra43}
\begin{equation}
\label{omega_poisson}
  \omega(r) = 4\pi n_0 r^2 \exp\left(-\frac{4}{3} \pi n_0 r^3\right).
\end{equation}
The average distance between a particle and its NN is given by 
\begin{equation}
\av{\Lambda} = \int_0^\infty r \omega(r) \, \dd r =
\left(\frac{3}{4 \pi n_0}\right)^{1/3} \Gamma_E
\left(\frac{4}{3}\right) \label{eq:Lambda_poisson}
\end{equation}
where $\Gamma_E$ is the Euler incomplete gamma function.

Let us now compute the probability, in a Poisson distribution, that
given a particle and its NN, they are mutually NN. Let us suppose that a
particle A has the particle B as NN at distance $r$. The probability
that A is the NN of B is equal to the probability that no other
particles are in the volume $v(r)$ defined by the portion of the
sphere of radius $r$ around $B$ which is not contained in the sphere
of radius $r$ around $A$. For a Poisson distribution this is
simply~\footnote{In a Poisson distribution, the probability that
  there are $k$ particles in a volume $V$ is given by $(n_0 V)^k
  \exp(-n_0 V)/k!$.}
\begin{equation}
p_2(r)=  \exp\left(-n_0 v(r)\right)
\end{equation}
where $v(r)$ is given by
\begin{equation}
v(r)=\frac{11}{12} \, \pi r^3 \;.
\end{equation}
Averaging on $r$, we get the probability 
that two particles are mutually NN:
\begin{equation}
\label{p_m}
p_3=\int_0^\infty \omega(r) \exp\left(-n_0 v(r)\right)\, \dd r \approx 0.6\; .
\end{equation}
Hence we have that more than the half of the particles are mutually NN.
If we restrict ourselves to particles which have a NN at a distance 
$l<\langle \Lambda \rangle$, this probability becomes
\begin{equation}
\label{p_m2}
p_4=\frac{\displaystyle{\int_0^{\av{\Lambda}} \omega(r) \exp\left(-n_0 v(r)\right)
  \, \dd r}}{\displaystyle{
\int_0^{\av{\Lambda}} \omega(s)\,\dd s}}
\approx 0.8\;.
\end{equation}
This result together with the fact that in a Poisson distribution the
force on a particle is mainly due to its NN \cite{chandra43},
allows us to consistently treat for an initial short time the many-body
problem as an ensemble of independent and isolated two-body systems.

\subsection{Time-scale of NN interaction}

This last result explains what happens if one leaves a Poisson
distribution without velocity evolving under its own gravity: most of 
particles will fall on their NN. Let us determine the time-scale of
this phenomenon.
To this aim, 
one can use conservation of energy in a pair of particles of
mass $m$:
\begin{equation}
E=- \frac{G m^2}{r_0} = \frac{m}{2} ( \dot{\ve{x}}_1^2 + \dot{\ve{x}}_2^2) -
\frac{G m^2}{r(t)}. \label{energy}
\end{equation}
where we have used the Newtonian potential.  As we will see in more
detail in the next subsection, the problem can be reduced to a single
dimension and choosing center of mass coordinates, we get $x_1(t) =
-x_2(t)$. After some algebraic
manipulations Eq. \eqref{energy} becomes
\begin{equation}
\dot{x}_1 = -\sqrt{Gm\left(\frac{1}{2x_1}-\frac{1}{r_0}\right)}
\end{equation}
assuming that $x_1(0)>0$. The time of fall is
\begin{equation}
\begin{split}
t_\text{fall}(r_0) &
= - \int_{r_0/2}^0
 \left[ G m\left( \frac{1}{2 x} -\frac{1}{r_0}
    \right)\right]^{-1/2}\, \dd x\\ &=
\frac{r_0^{3/2} \pi}{4}\frac{1}{\sqrt{G m}}
\label{tfall}
\end{split}
\end{equation}
Taking  for $r_0$ the mean distance between NNs, $\av{\Lambda}$ given
by Eq.~\eqref{eq:Lambda_poisson}, we get 
\be
\label{tau}
\tau 
= \frac{\pi}{4}\sqrt{3 \Gamma_E^3 (4/3)}\,
\frac{1}{\sqrt{4\pi G \rho_0}}
\approx
 \frac{1.148}{\sqrt{4\pi G \rho_0}}
\ee
where $\rho_0=m n_0$ is the mass density.

  \subsection{Approximate evolution of the conditional average density}

As already mentioned,
the force on a particle in a Poisson distribution is almost only due
to its NN. As in our simulations the particles have no initial
velocity, they will start to fall in direction of their NN and we will
see that this is what explain the early evolution of $\av{
n(\ve{r})}_p$ for a time $t\lesssim \tau$.

  Let us consider that the interaction potential is
  $U(\ve{r})=U(r)$~\footnote{We do not restrict ourselves to a
  precise potential as it can vary in different NBS.}.
  As said before, we make the assumption that the force on a
  particle is only due to its NN and that the
  Poisson distribution can be approximated by an ensemble of
  particle pairs evolving independently.
  The evolution of one of these pairs is given by the following equations:
  \begin{subequations}  
    \begin{align}
      m \ddot{\ve{x}}_1 &= -\nabla_{\ve{x}_1} U(r_{12}) =
      -\frac{\dd U}{\dd r}\biggr|_{r_{12}}\cdot
      \frac{\ve{x}_1-\ve{x}_2}{r_{12}} \\
      m \ddot{\ve{x}}_2 &= -\nabla_{\ve{x}_2} U(r_{12}) =
      +\frac{\dd U}{\dd r}\biggr|_{r_{12}}\cdot
      \frac{\ve{x}_1-\ve{x}_2}{r_{12}}
    \end{align}
  \end{subequations}
  with $r_{12}=|\ve{x}_1-\ve{x}_2|$. Adding these two equations, one gets
  $\ddot{\ve{x}}_1 = -\ddot{\ve{x}}_2$ (conservation of total momentum). As
  the particles are supposed to be at rest at $t=0$, one has $\ve{x}_1 =
  -\ve{x}_2$ with a proper choice of the origin (center of mass coordinates).
  With this relation,
  $\ve{x}_1-\ve{x}_2=2\ve{x}_1=-2\ve{x}_2$ and one has to solve only one
  equation of motion, for particle 1 for instance:
  \begin{equation}
    m \ddot{\ve{x}}_1 =
    -\frac{\dd U}{\dd r}\biggr|_{2|\ve{x}_1|}\cdot \frac{\ve{x}_1}{|\ve{x}_1|}.
  \end{equation}
  Using again the fact that the initial velocity is null, one can reduce
  the number of dimensions to one:
  \begin{equation}
    m \ddot{x} =
    -\frac{\dd U}{\dd r}\biggr|_{2|x|}\cdot \mathrm{sign}(x) =
    -\frac{\dd V}{\dd x} \label{eq:motion}
  \end{equation}
  with $V(x) = U(|2x|)/2$ which is the equation for the evolution of a
  single particle in the potential $V$. One can now use Liouville
  theorem \cite{dorfman}  in order to study
  the evolution of a phase space density function of systems evolving
  according to this equation and choosing an appropriate density
  function, one can obtain the evolution of $\omega(r)$.

  If $f(x,v,t)$ is a phase space density function, Liouville theorem
  states that
  \begin{equation}
    (\partial_t + v \partial_x + \dot{v} \partial_v) f = 0 \; ,
  \end{equation}
  where, in our case, $m\dot{v}= -\dd V/\dd x$. The appropriate initial
  condition is
  \begin{equation}
    f(x,v,0) = \delta(v) \frac{\omega(2|x|)}{2} \label{eq:init}.
  \end{equation}
  with $\omega(r)$ given by \eqref{omega_poisson}.
  We divide it by 2 in order to have half of the particles with $x>0$
  and half with $x<0$. Knowing $f(x,v,t)$, one can obtain $\omega(r,t)$,
  the time evolved NN probability distribution, with
  \begin{equation}
    \omega(r,t) =\int_{-\infty}^\infty f(-r/2,v,t)\, \mathrm{d}v + \int_{-\infty}^\infty f(r/2,v,t)\, \mathrm{d}v .
    \label{eq:omegart}
  \end{equation}

  In order to solve the Liouville equation, let us
  denote by $\phi_t(x_0,v_0) =(X_t(x_0,v_0),V_t(t,x_0,v_0))$ the solution of the
  equation of motion with initial condition $x_0,\ v_0$ at $t=0$.
  The Liouville equation implies 
  that $f(x,v,t)$ remains constant along a phase space trajectory:
  \begin{equation}
    f\left(X_t(x_0,v_0),V_t(x_0,v_0),t\right) 
    = f(x_0,v_0,0).
  \end{equation}
  With our initial conditions, the solution of this equation is
  therefore 
  \begin{equation}
    \begin{split} \label{eq:f_sol}
      f(x,v,t) &=  f(\phi_{-t}(x,v),0) \\
      &= \iint_{\mathbb{R}^2} \dd x_1 
      \dd v_1 \ \left[ f(\phi_{-t}(x_1,v_1),0) \right. \\ 
	& \left. \qquad   \times \ \delta(x-x_1)\  \delta(v-v_1)  \right]\\
      &= \iint_{\mathbb{R}^2} \mathrm{d}x_0 
      \mathrm{d}v_0 \ \left[ f(x_0,v_0,0) \right. \\ 
	& \left. \qquad
	\times \ \delta(x-X_t(x_0,v_0))\  \delta(v-V_t(x_0,v_0))  \right]
    \end{split}
  \end{equation}
  with $f(X_t(x,v),V_t(x,v),t)\equiv f(\phi_t(x,v),t)$.  We have used
  the fact that the determinant of the Jacobian matrix in the change
  of variables from $(x_1,v_1)$ to $(x_0,v_0)$ is $\det(\partial
  \phi_t /\partial(x,v))=1$.  This is actually related to the
  Liouville theorem~\cite{gallavotti}.

  With this solution, we can get the evolution of $\omega(r,t)$.
  First let us remark that $f(x,v,0)=f(-x,-v,0)$ and as the force in
  (\ref{eq:motion}) is odd, if $x(t)$ is a solution, $-x(t)$ is also a
  solution. This permits to show that $f(x,v,t)=f(-x,-v,t)$.  It is
  then easy to see that (\ref{eq:omegart}) can be rewritten as
  \begin{equation} \omega(r,t) =2 \int_{-\infty}^\infty f(r/2,v,t)\,
  \mathrm{d}v .  \label{eq:omegart2} \end{equation}
Using this last equation and Eq. \eqref{eq:f_sol}, one has:
\begin{multline}
\omega(r,t)=2\int_{-\infty}^\infty \mathrm{d}x_0
\int_{-\infty}^\infty \mathrm{d}v_0 \, f(x_0,v_0,0)\  \\
\times \delta(r/2-X_t(x_0,v_0)).
\end{multline}
As $f(x,v,0)=\delta(v)\omega(2|x|)/2$, this becomes
\begin{equation}
\omega(r,t) = \int_{-\infty}^\infty \mathrm{d}x_0 \ \omega(2|x_0|)\
\delta(r/2-X_t(x_0,0)).
\end{equation}
Using the fact that 
for a function $f:\mathbb{R}\to\mathbb{R}$  one has
\begin{equation}
\delta(f(x)) = \sum_{y\in Z(f)} \frac{\delta(x-y)}{|f'(y)|}
\end{equation}
with $Z(f)=\{y\in\mathbb{R}\ | \ f(y)=0\}$, we get
\begin{equation}
\omega(r,t) = \sum_{x_0\in S_{t}^r}
\left| \frac{\mathrm{d}X_t(x_0,0)}{\mathrm{d}x_0} \right|^{-1} \omega(2|x_0|)
\label{omega_analytic}
\end{equation}
with $S_t^r = \{ x_0 \in \mathbb{R}\ | \ X_t(x_0,0) = r/2\}.$ Of
course there are points $x_0$ in $S_t^r$ such that
$\dd X_t(x_0,0)/\dd x_0=0$ and therefore $\omega(r,t)$ is not well
defined at some isolated points.

We may solve numerically Eq. \eqref{eq:motion} to find a solution for
$X_t(x_0,0)$. The steps to get $\omega(r,t)$ for a given $t$ are the
following.  We start with a set $\{x_{0,i} = x_{0,\text{min}}+i \delta
\ | \ \delta>0,\ 0\leq i \leq n\}$ where $x_\text{min}$, $\delta$ and
$n$ are chosen so that the region covered gives non-negligible values
for $\omega(2x)$ and that this region is sufficiently sampled. For
each $i$, one calculates numerically $X_i\equiv X_t(x_{0,i},0)$. By
doing a linear interpolation with these values, we have an estimate of
$X_t(x_0,0)$ for all $x_0$ in the region covered by the $x_{0,i}$.
The last step is to find the set of $x$ which solve $X_t(x,0)=r/2$.
Once we have $\omega(r,t)$, we find the conditional density by using
Eq.~\eqref{omega1}.

\subsection{Comparison with a simulation}

In order to test the simple argument presented in the last subsection,
we did a N-body simulation. We have used the code
\textsc{Gadget}~\cite{gadget} based on a tree algorithm. The infinite
universe is simulated by using periodic boundary conditions and the
usual Ewald summation technique. The force between two particles is
not exactly Newtonian but a softened one is used~\cite{binney}.  Note
that the potential used is not the standard Plummer one but a similar
one which has the advantage of being perfectly Newtonian at a scale
larger than the softening length.

We have generated a Poisson distribution with $N=32^3$ particles in a
box of nominal size $L$. The mass of the particles is such that the
mass density is one.  The softening length is $\epsilon = 0.00175 L$:
by using Eq. \eqref{eq:Lambda_poisson} we find $\av{\Lambda}\approx
0.017L$ and hence $\eta \approx 10$.  The initial velocities are set
to zero, and the simulation is run up to $4\, \tau$.

\begin{figure}
\includegraphics[width=9cm,angle=0]{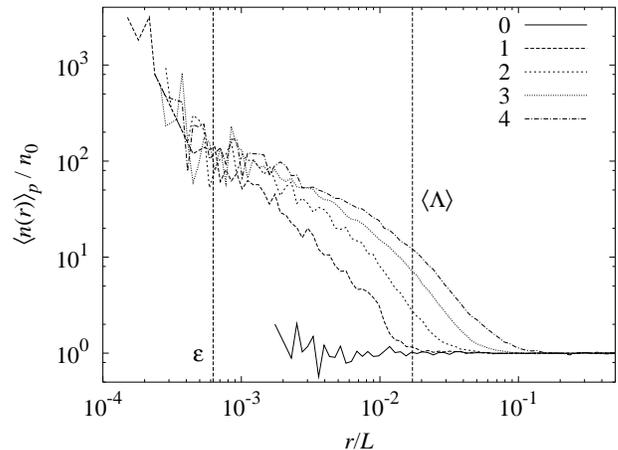}
\caption{The normalized conditional density in a Poisson
distribution at time $\tau,\ 2\,\tau,\ 3\,\tau,\ 4\,\tau$.
Note that once correlations are developed, the subsequent evolution
increases the range of scales where non-linear 
($ \langle n(r) \rangle_p \gg n_0$) 
clustering  is formed, while the function behavior of two-points 
remain unchanged. 
\label{gamma_poisson}}
\end{figure}

The time evolution of the conditional density is shown in
Fig.~\ref{gamma_poisson} (here and in what follows we normalize the
conditional average density to the asymptotic density, i.e. we
consider $\av{n(r)}_p / n_0$).  It is worth noticing that once the
power-law correlations are developed, the subsequent evolution
increases the range of scales where non-linear clustering is formed,
i.e.  where $\av{ n(r) }_p \gg n_0$, by approximatively a
simple rescaling: denoting by $\av{n(r,t)}_p$ the conditional density
at time $t$, one has
\begin{equation}
\av{n(r,t+\delta)}_p \approx \av{n(a\cdot r,t)}_p 
\label{eq:rescaling}
\end{equation}
where $a>0$ is a constant which depends on the time~\cite{slbj03}.

\begin{figure}
\includegraphics[width=9cm,angle=0]{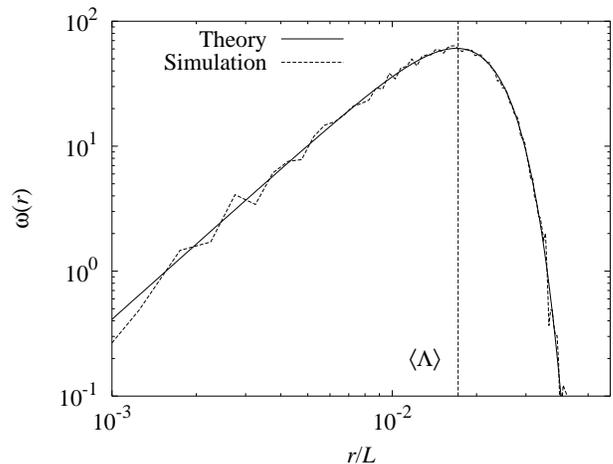}
\caption{Initial NN probability distribution for the Poisson case.
The solid line is the exact solution given by
Eq. \eqref{omega_poisson}
while the dashed line is the measured one in the simulation.
\label{omega_poisson_0}}
\end{figure}

\begin{figure}
\includegraphics[width=9cm,angle=0]{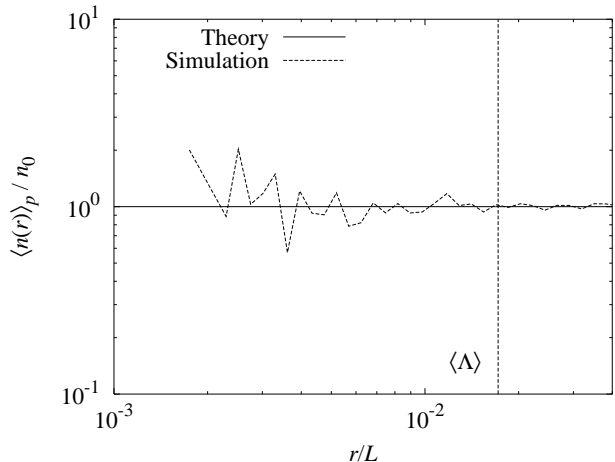}
\caption{
Normalized two-point conditional density at the initial time:
the solid line (Theory) is the theoretical ensemble average
behavior while the dashed line is the measured one in the
simulation.
\label{gamma_poisson_0}  }
\end{figure}

In Fig.~\ref{omega_poisson_0} we show the initial NN density
distribution obtained from the Poisson distribution used in the
simulation and the one from Eq.~\eqref{omega_poisson}.  The
conditional density of the initial configuration together with the one
obtained by using Eq.~\eqref{omega1} are shown in
Fig.~\ref{gamma_poisson_0}.


\begin{figure}
\includegraphics[width=9cm,angle=0]{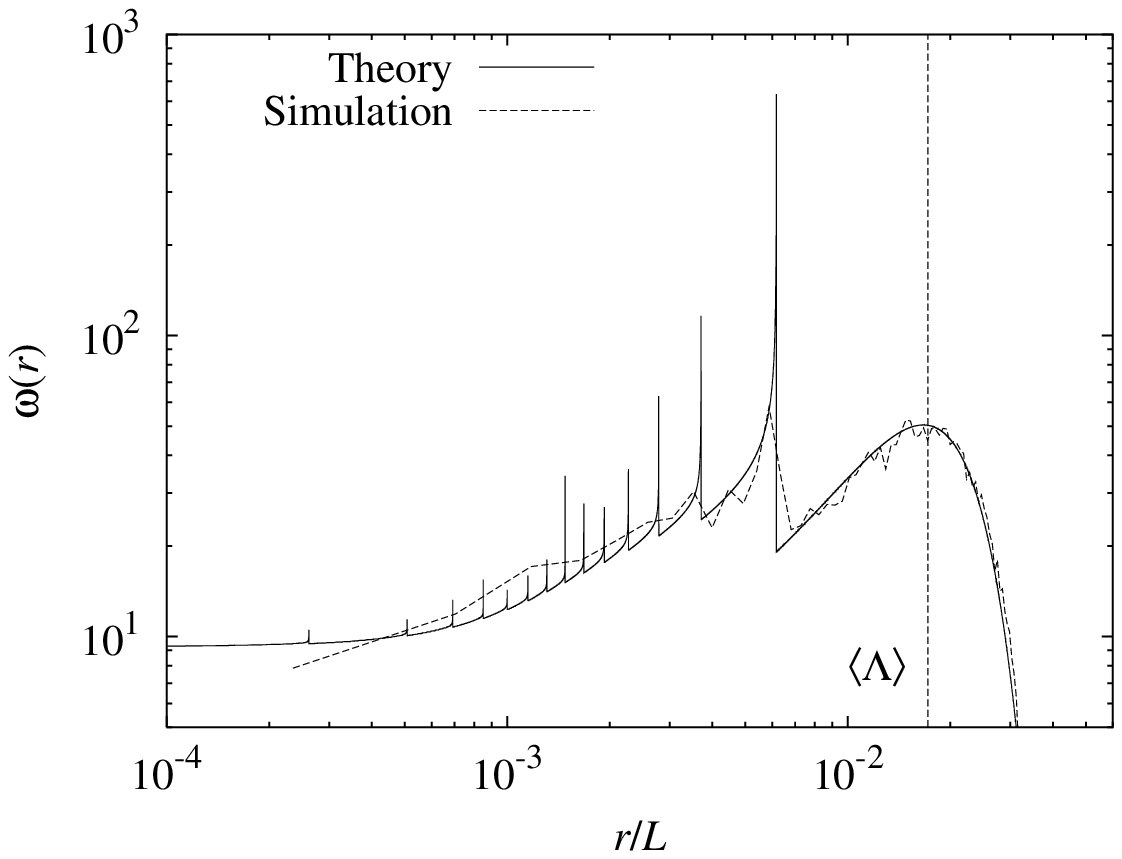}
\caption{NN probability distribution 
at $t=0.5 \,\tau$ in the Poisson simulation.
\label{omega_poisson_5}}
\end{figure}

\begin{figure}
\includegraphics[width=9cm,angle=0]{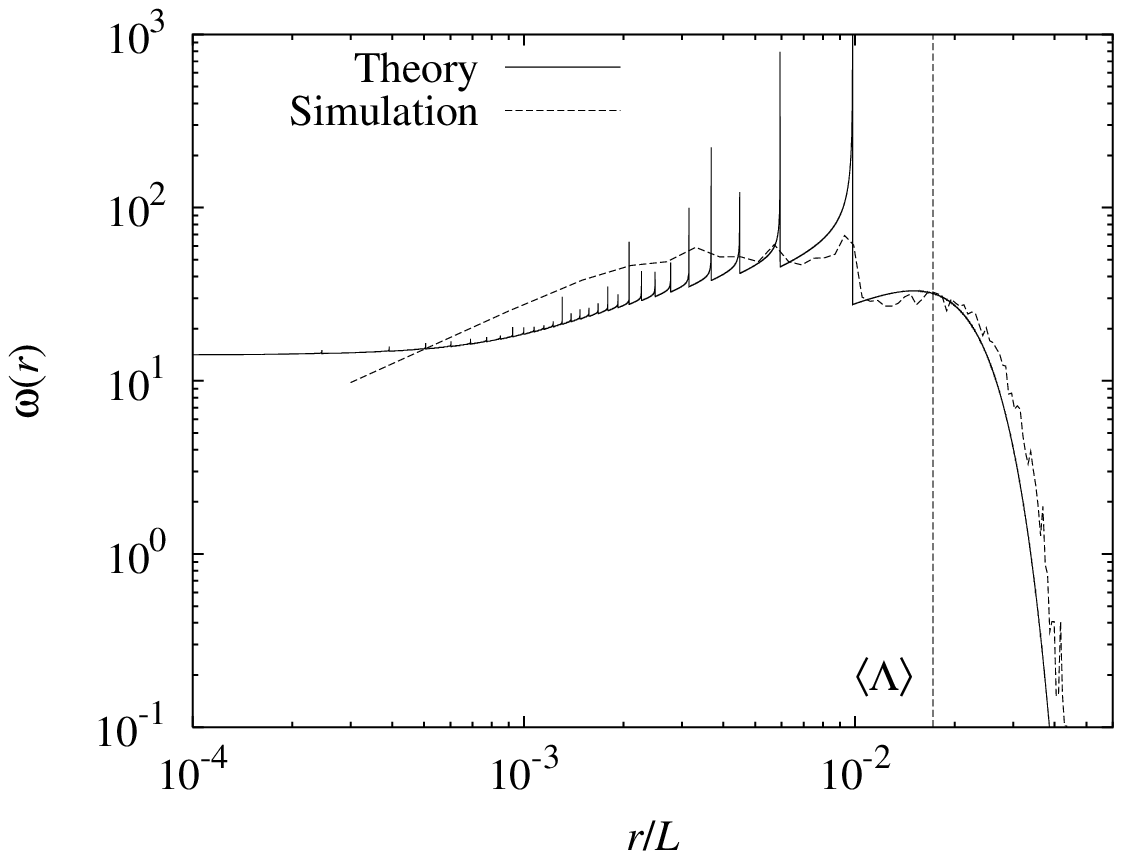}
\caption{NN probability distribution at $t=\tau$ in the Poisson simulation.
\label{omega_poisson_10}}
\end{figure}

\begin{figure}
\includegraphics[width=9cm,angle=0]{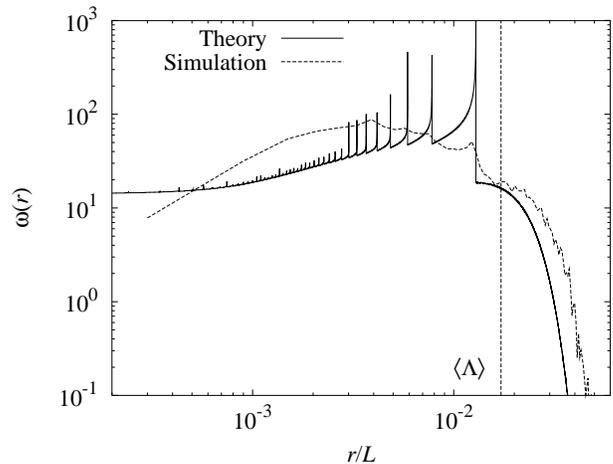}
\caption{NN probability distribution at $t=1.5\,\tau$ in 
the Poisson simulation.
\label{omega_poisson_15}}
\end{figure}

The evolution of the NN probability distribution in the simulation
together with the one obtained from \eqref{omega_analytic}, at times
$0.5\, \tau$, $\tau$ and $1.5\, \tau$, is shown in
Figs~\ref{omega_poisson_5}-\ref{omega_poisson_15}.  We may notice that
the agreement is quite good and even excellent at $t=0.5\,\tau$.  The
differences which appear at $t=\tau$ and $t=1.5\,\tau$ seem to be
explained by the following arguments.

First of all we remind that in a Poisson distribution
the force acting on a particle can be decomposed 
in two terms: the one given by the NN particle and 
the one due to all the other particles. While the first 
represents a large contribution, the second 
rapidly goes to zero for symmetry reasons \cite{chandra43}. 
However, for particles which have a NN further than the average $\LL$,
the situation is different. Let us denote by $A$ such a particle and
its NN by $B$. The force $F_{BA}$ from the latter on $A$ being weaker
than the average force on a particle from its NN, the force
contribution of other particles nearby becomes also important on the
total force on $A$. This total force is then not necessarily in the
direction of $B$ and the particle $A$ will not ``fall'' on
it. Furthermore the particle $B$ has a high probability of having a NN
different from $A$; it should therefore not go towards $A$. The
distance between $A$ and its NN $B$ should then grow. This is actually what
we observe if we compare carefully Figs~\ref{omega_poisson_0} and
\ref{omega_poisson_15}: looking the value of $r/L$ at large scales at
which the NN probability distribution reach a value of $10^{-1}$, we
see that it is $3\cdot 10^{-2}$ at $t=0$ and $3.5\cdot 10^{-2}$ at
$t=1.5\,\tau$, i.e.  the particles whose NN is at a distance
$3\cdot 10^{-2}L$ initially are at a larger distance ($3.5\cdot
10^{-2}L$) at $t=1.5\,\tau$.

Secondly, concerning particles which have their NN at a distance
closer than the average $\LL$ we observe that at scales between
$10^{-3}L$ and $5\cdot 10^{-3}L$ a bump is created: our simple model
predicts less particles than observed in the simulation. This seems to be a
sign of the creation of larger structures. If two particles are
isolated, they will move in a regular oscillating motion.  This is
what the model predicts. In the simulation these two particles,
i.e. a particle and its NN, will move together for a while as in
the model but in the same time be attracted toward another pair or group of
particles, which is not described by the model. This could have the
effect of bringing the two particles closer together and even give
rise to an exchange of NN with the 
other group of particles, making the evolution of
the NN probability distribution evolving differently from the
model. The bump reflects therefore this step of the clustering which
tends to bring pairs together.


\begin{figure}
\includegraphics[width=9cm,angle=0]{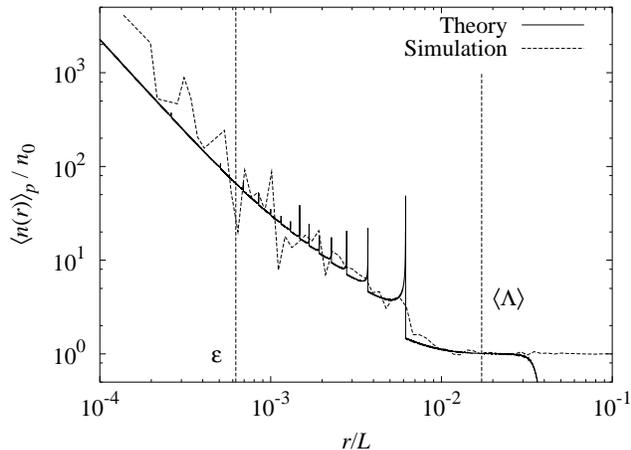}
\caption{Behavior of the average
conditional density at $t =0.5 \,\tau$ in the Poisson simulation. \label{gamma_poisson_5}}
\end{figure}

\begin{figure}
\includegraphics[width=9cm,angle=0]{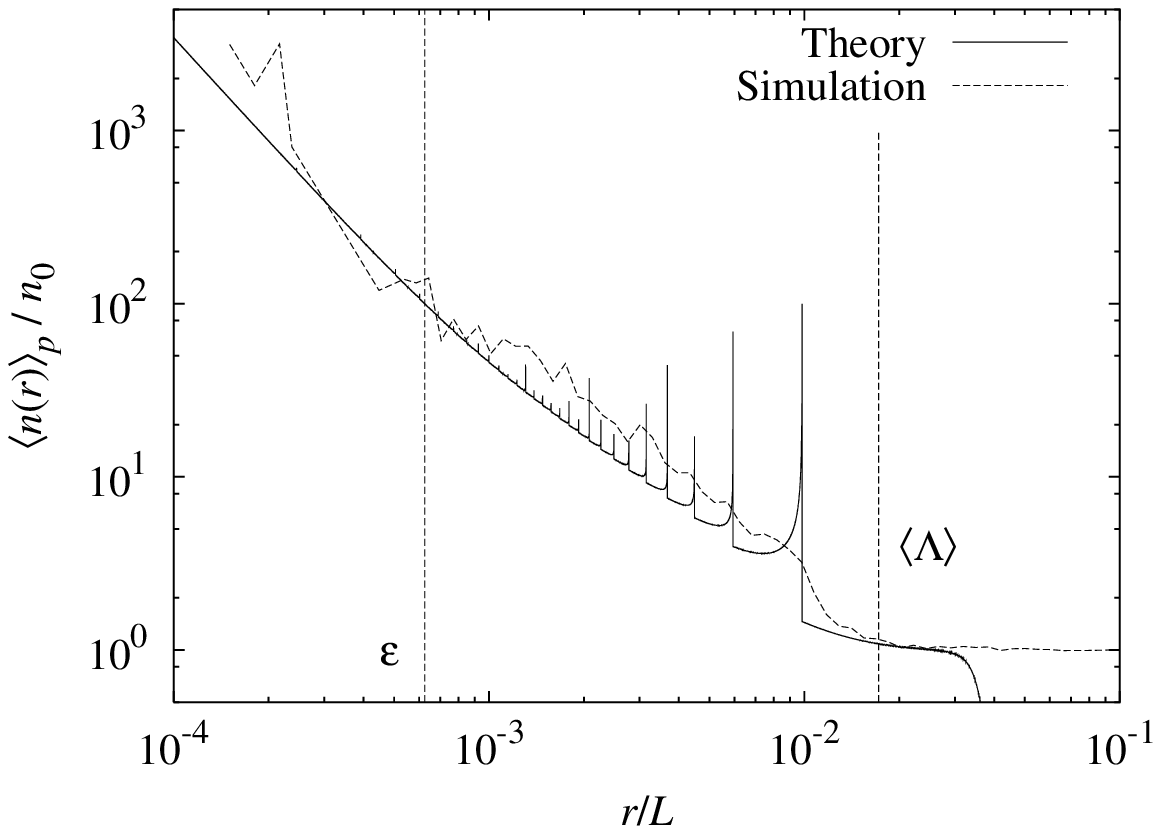}
\caption{Behavior of the average
conditional density at $t =\tau$ in the Poisson simulation. \label{gamma_poisson_10} }
\end{figure}

\begin{figure}
\includegraphics[width=9cm,angle=0]{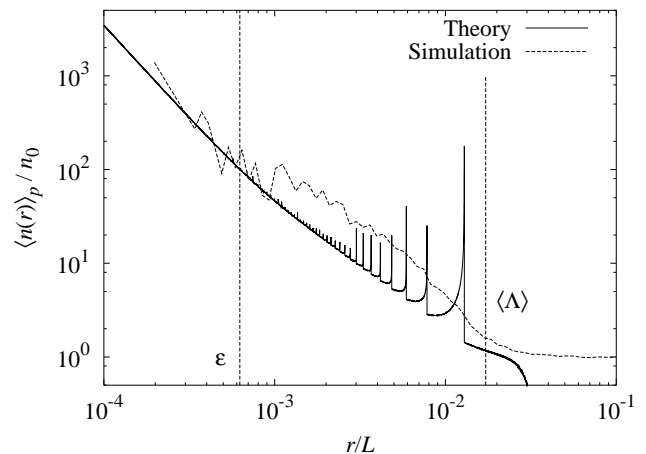}
\caption{Behavior of the average
conditional density at $t =1.5 \,\tau$ in the Poisson 
simulation.\label{gamma_poisson_15} }
\end{figure}

In Figs~\ref{gamma_poisson_5}-\ref{gamma_poisson_15} we compare the
predictions of the conditional density given by our model, with the
measured ones from the simulations at times $0.5\,\tau$, $\tau$ and
$1.5\,\tau$.  One sees that our approximation works again really well
as it succeeds in reproducing the development of the
correlations. This means that these correlations are therefore only a
consequence of the interaction of NNs. We may also notice an
interesting thing: at $t=1.5\,\tau$, even if the agreement is
marginally good at scales larger than $10^{-3}L$, it is still correct
at smaller scales. An explanation is that these scales correspond to
pairs whose particles were very close (i.e. $< \LL$)
initially and therefore well bounded. When they start to feel
the effect of particles around, their relative motion is not affected
and is still described by a two-body interaction.

At larger scales, where there are no correlations, our approximation
fails to reproduce the correct behavior at all times. For a certain
$r$ the conditional density goes rapidly to 0. This is due to the
fact that at these scales, the NN probability distribution is really
small and the Eq.~\eqref{omega1} is not valid anymore: this
equation implies that the density around a particle is only due to its NN
and that there are no particles further than the NN. Therefore at
distances larger than the average distance between NNs, the density has
to go to 0 as there are no other particles to maintain a non-zero
density.

%


We finally remark that, as noticed in \cite{slbj03}, to study the role
of these NN interactions in the evolution of clustering, one may
modify the force integrator of the numerical code to include {\it
only} the NN contribution to the gravitational force. Of course, the result
agrees perfectly with the study presented here.

\subsection{Poisson with large softening}

In the Poisson simulation, we have observed that the first structures
created are pairs of particles. Now we present another simulation in
which this is not the case.  It is simply a Poisson simulation with a
large softening, one hundred times larger than in the previous case:
$\epsilon=0.175L$ and hence $\eta \approx 0.1$.

\begin{figure}
  \includegraphics[width=9cm,angle=0]{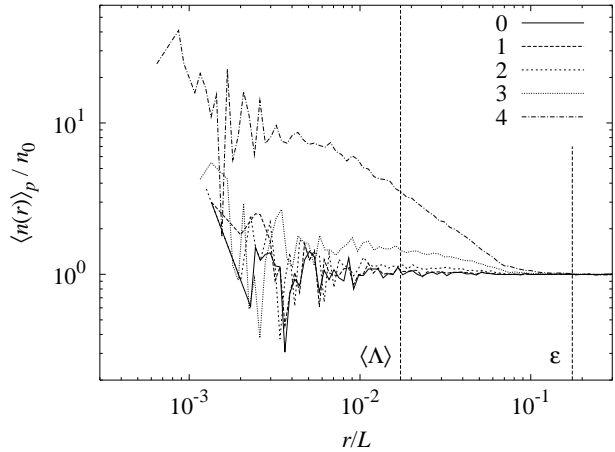}
  \caption{Evolution of the normalized conditional density
    in the Poisson with large softening simulation.
    The times are $0,1,2,3,4$ in units of $\tau$. \label{fig:P32_1_gamma} }
\end{figure}

Figure \ref{fig:P32_1_gamma} shows the evolution of the conditional
density in this simulation. The time is still in unit of $\tau$ but
only for comparison with the first Poisson simulation because this is
not anymore a microscopic characteristic time.  One can see that the
correlations do not develop at the smallest scales of the system
($\av{\Lambda}=0.017L$) but are directly found up to $10^{-1}L$ which
is of the order of $\epsilon$.

\begin{figure}
  \includegraphics[width=9cm,angle=0]{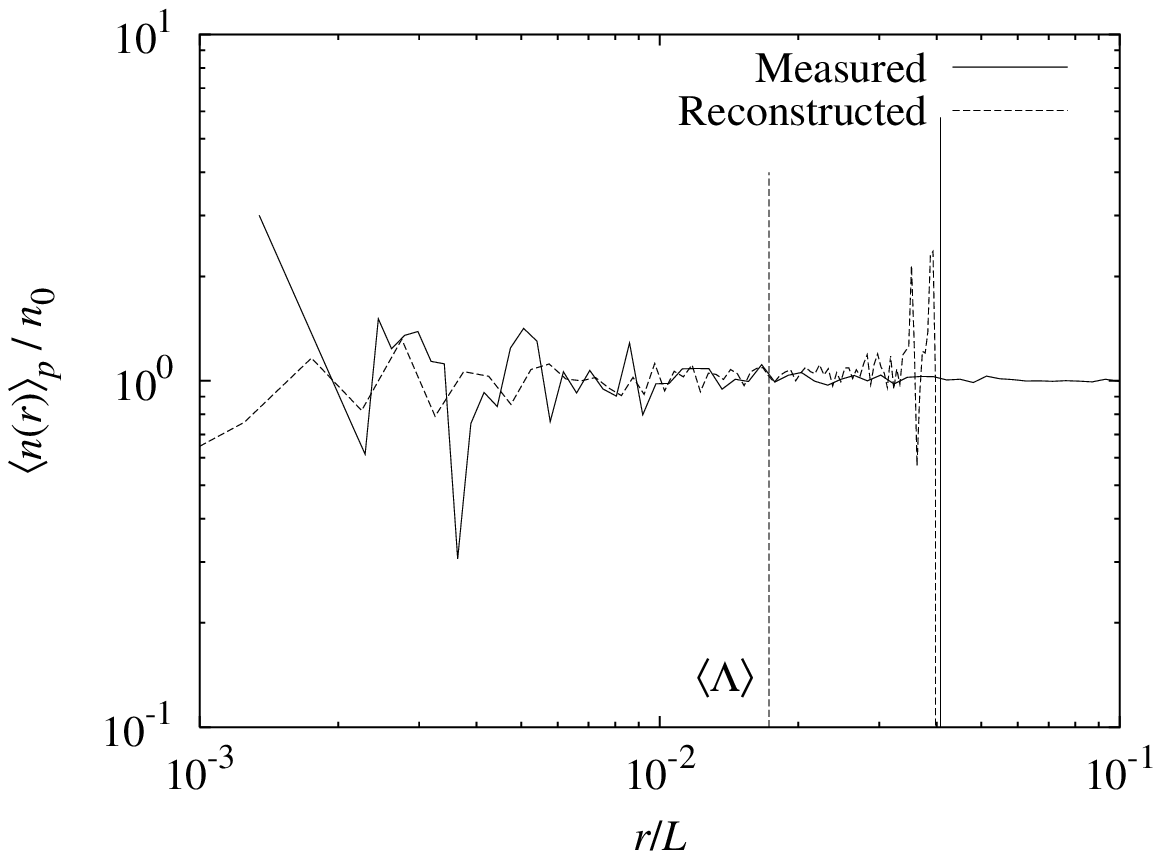}
  \caption{Reconstruction of the conditional density from $\omega(r)$
  at $t=0$ in the Poisson with large softening simulation.
    \label{omega_gamma_P32_1.0_0}}
\end{figure}

\begin{figure}
  \includegraphics[width=9cm,angle=0]{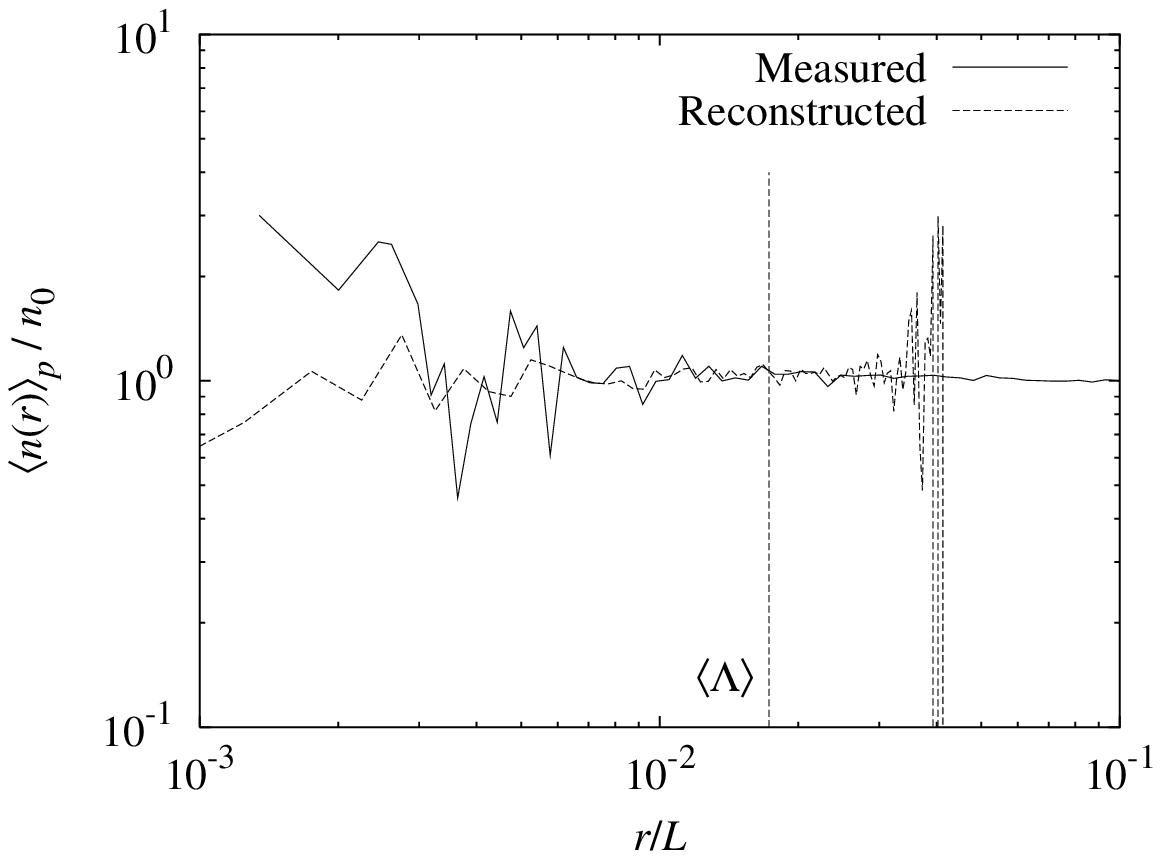}
  \caption{Reconstruction of the conditional density from $\omega(r)$
    at $t=\tau$
in the Poisson with large softening simulation.
    \label{omega_gamma_P32_1.0_10}}
\end{figure}

\begin{figure}
  \includegraphics[width=9cm,angle=0]{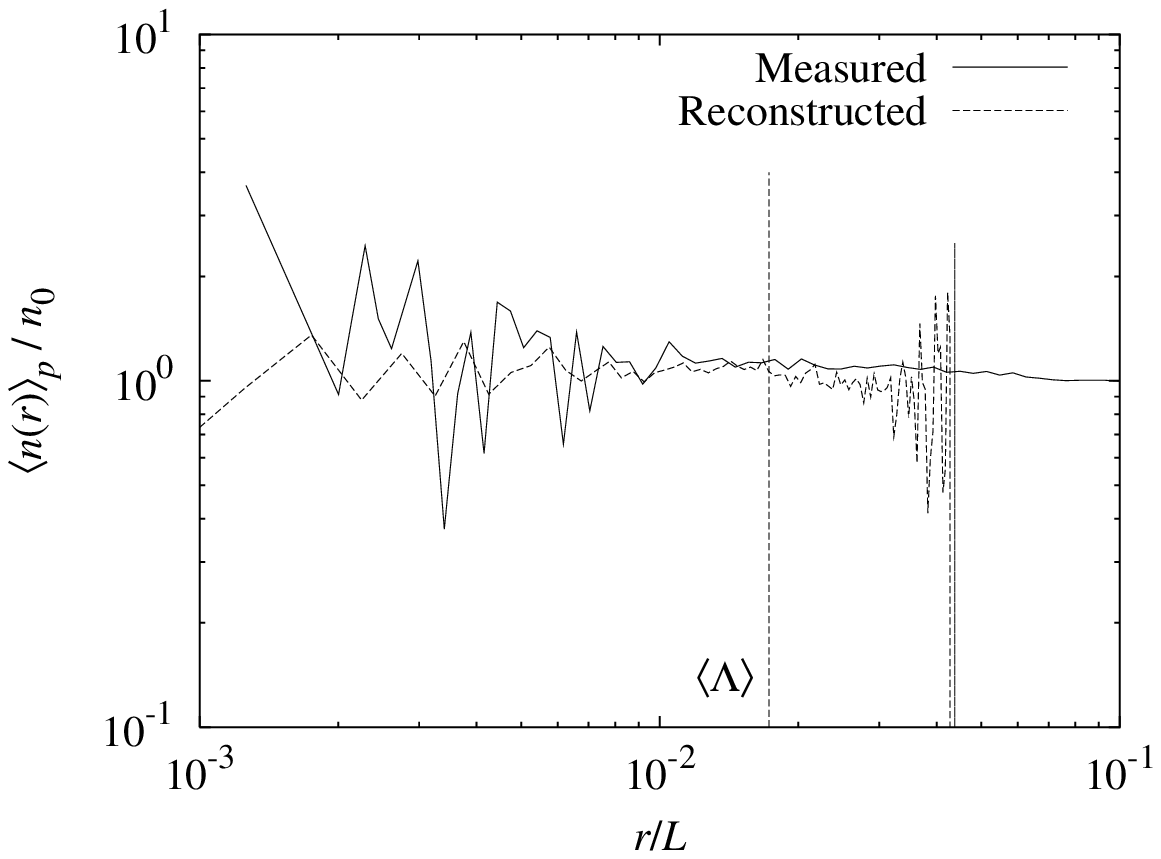}
  \caption{Reconstruction of the conditional density from $\omega(r)$
    at $t=2\,\tau$
in the Poisson with large softening simulation.
    \label{omega_gamma_P32_1.0_20}}
\end{figure}

\begin{figure}
  \includegraphics[width=9cm,angle=0]{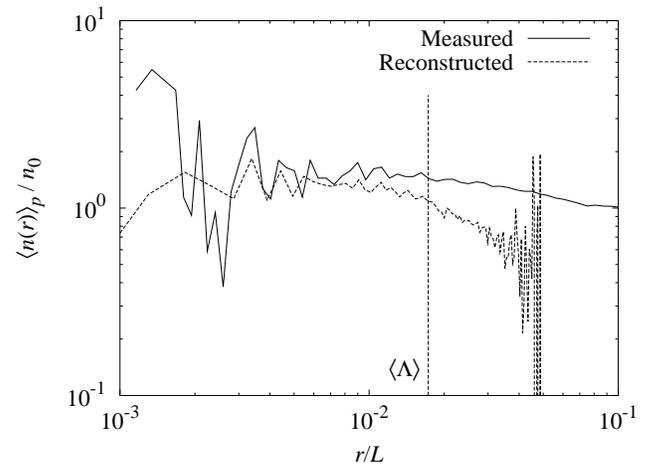}
  \caption{Reconstruction of the conditional density from $\omega(r)$
    at $t=3\,\tau$
in the Poisson with large softening simulation.
    \label{omega_gamma_P32_1.0_30}}
\end{figure}

Looking now at
Figs~\ref{omega_gamma_P32_1.0_0}-\ref{omega_gamma_P32_1.0_30}, where
we compare the conditional densities obtained from the simulation and
the ones reconstructed from the NN probability distributions, we see
that as soon as correlations develop they are already made of many
particles as the approximations of the conditional density by the NN
probability distribution fails.

In the beginning of this simulation, the NN contribution to the total
force acting on a particle is clearly not important. The dominant
contribution is actually the force due to many particles at some
larger scales.  This means that two nearby particles do not fall on
each other as in the previous case but feel approximatively the same
force and therefore go in the same direction once the simulation
starts. This direction should be the one of the nearest mass
over-density. Some other particles will also be attracted in this
direction. The effect of this motion is the formation of the first
structures, directly made of more than two particles.


As a final remark, it is interesting to note that when power-law
correlation are formed at $t\approx 4 \tau$ the exponent and the
amplitude of the conditional density agrees very well with the
simulation with small softening previously discussed.


\section{The shuffled lattice and the CDM case} 

We study now two different cases where the average force on a particle
in the initial distribution is different from the Poisson case,
i.e. it is not dominated by the NN one. The main point here is
to use the relation \eqref{omega1} to study the creation of the first
structures: by obtaining the NN probability distribution in a
simulation, we reconstruct the conditional density and compare it with
the one measured directly in the simulation.

\subsection{The shuffled lattice}

A shuffled lattice is a simple ordered distribution \cite{GJS} which
is obtained by adding a random small perturbation to a perfect lattice
of particles: each particle of this lattice is moved randomly in a
cubic box centered on the unperturbed position of the particle.  The
only parameter is then the ratio 
\begin{equation}
 a_\text{s} = \frac{\delta}{l} 
\end{equation}
between the size of the cubic box $2\delta$ and the lattice spacing
$l$.  When $a_\text{s}=0$, it is a perfect lattice while as $a_s
\rightarrow \infty$ it becomes a Poisson distribution \cite{GJS}.  For
the simulation presented here, we have used a shuffled lattice with
$32^3$ particles, and shuffling parameter $a_\text{s}=0.25$.  The mass
of the particles, the number density and the softening length of the
force are the same as for the Poisson simulations previously
discussed: this gives $\eta \approx 14$.

\begin{figure}
\includegraphics[width=9cm,angle=0]{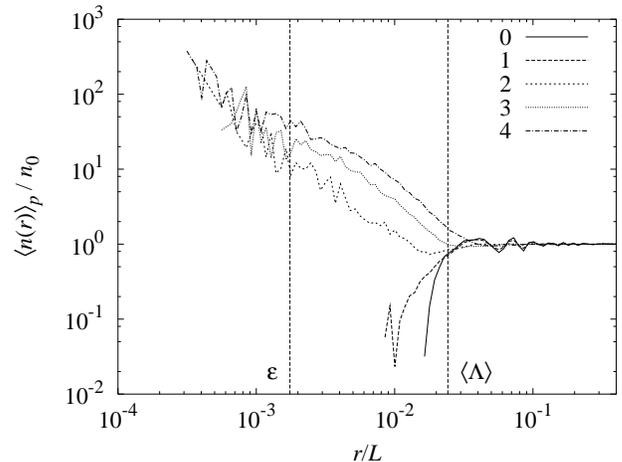}
\caption{Evolution of the normalized conditional density
in the shuffled lattice distribution.
The times are $0,1,2,3,4$ in units of $\tau$.
\label{gamma_sl}}
\end{figure}

In Fig.~\ref{gamma_sl}, the evolution of the conditional density is
shown. The time goes from $0$ to $4\,\tau$ with $\tau$ given by
Eq.~\eqref{tau}. One may note that once correlations are developed, the
evolution proceeds in a very similar way to the Poisson case (see
Fig.~\ref{gamma_poisson}): it is the same kind of rescaling as given by
Eq.~\eqref{eq:rescaling}, the only difference being the speed at which
this happens~\cite{slbj03}.

\begin{figure}
\includegraphics[width=9cm,angle=0]{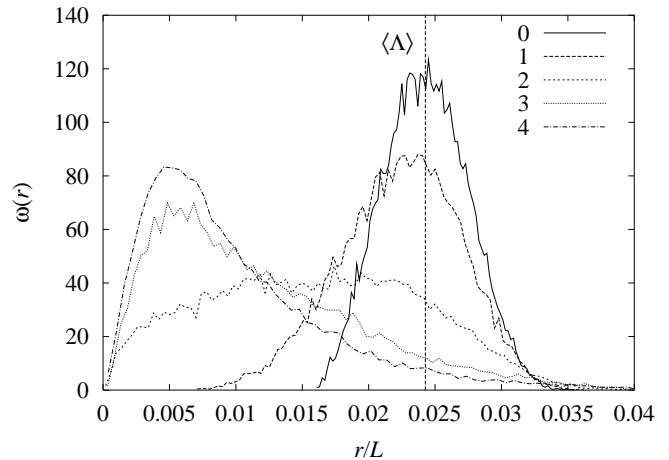}
\caption{Evolution of the NN probability distribution in the shuffled lattice for the same
  times than in Fig.~\ref{gamma_sl}
\label{omega_sl}}
\end{figure}

In Fig.~\ref{omega_sl}, it is shown the NN probability distribution
measured in the simulation at the corresponding times.  It is
important to note that at $t=0$ there is an anti-correlation at small
scales: the normalized conditional density is smaller than 1. This is
due to the fact that two particles cannot be closer than a minimal
distance which depends on $a_\text{s}$: the excluded volume feature being a
typical property of \emph{super-homogeneous} distributions \cite{GJS}.  This
can be seen with the NN probability distribution which is very
peaked around the mean inter-particle separation.

\begin{figure}
\includegraphics[width=9cm,angle=0]{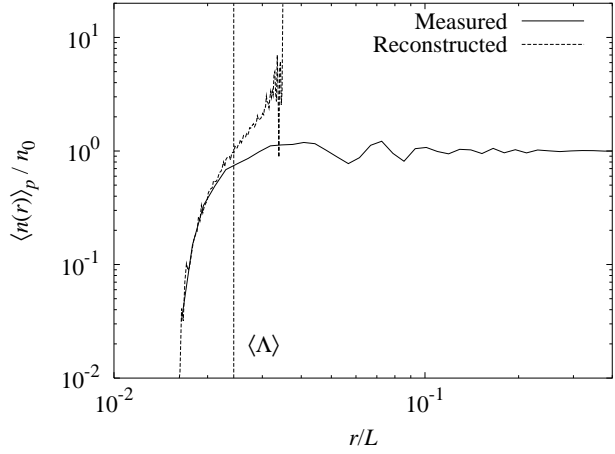}
\caption{Reconstruction of the conditional density from $\omega(r)$ at $t=0$ in the shuffled lattice.
\label{omega_gamma_sl_0}}
\end{figure}

\begin{figure}
\includegraphics[width=9cm,angle=0]{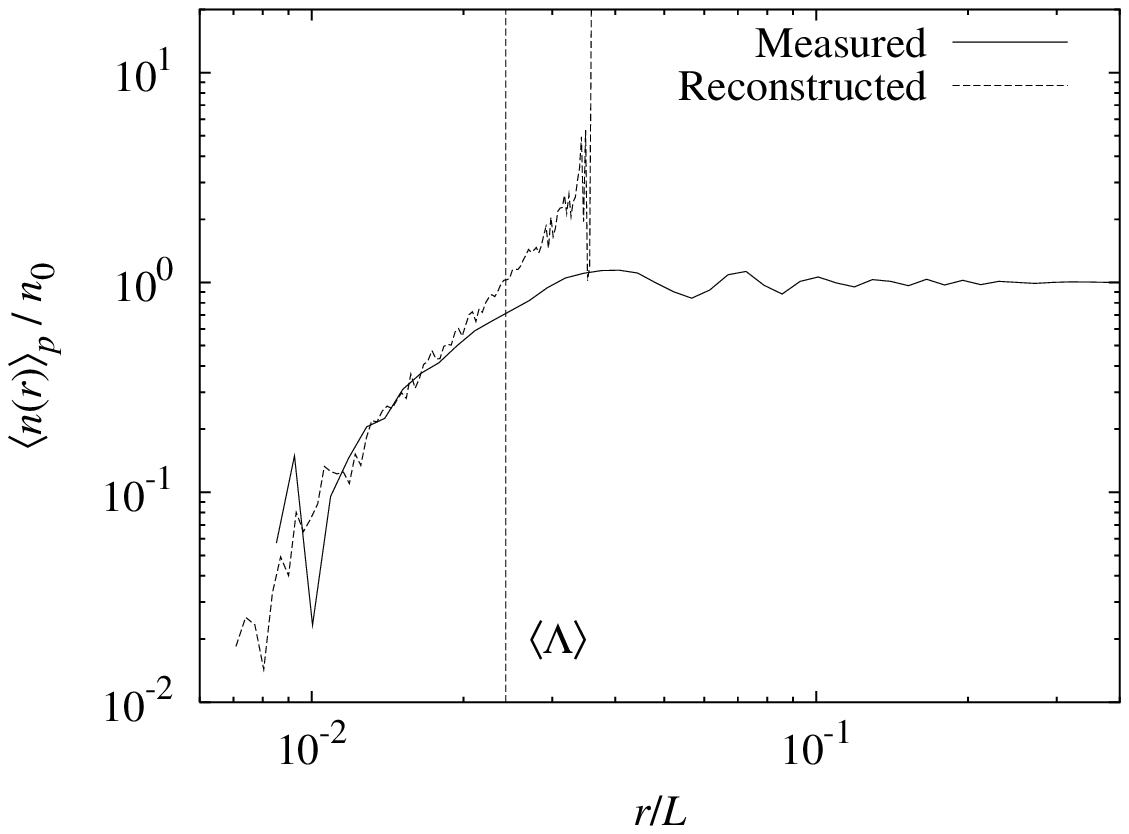}
\caption{Reconstruction of the conditional density from $\omega(r)$ at $t=\tau$ in the shuffled lattice.
\label{omega_gamma_sl_10}}
\end{figure}

\begin{figure}
\includegraphics[width=9cm,angle=0]{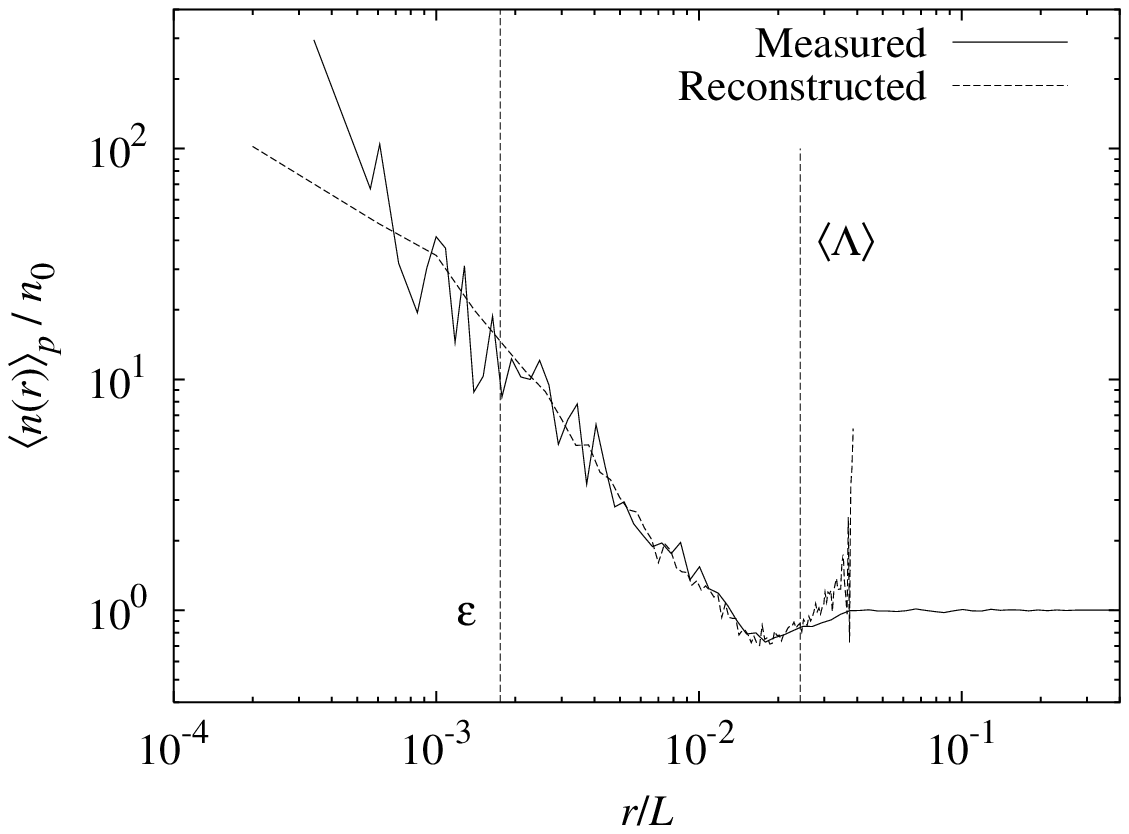}
\caption{Reconstruction of the 
conditional density from $\omega(r)$ at $t=2\,\tau$ in the shuffled lattice.
\label{omega_gamma_sl_20}}
\end{figure}

\begin{figure}
\includegraphics[width=9cm,angle=0]{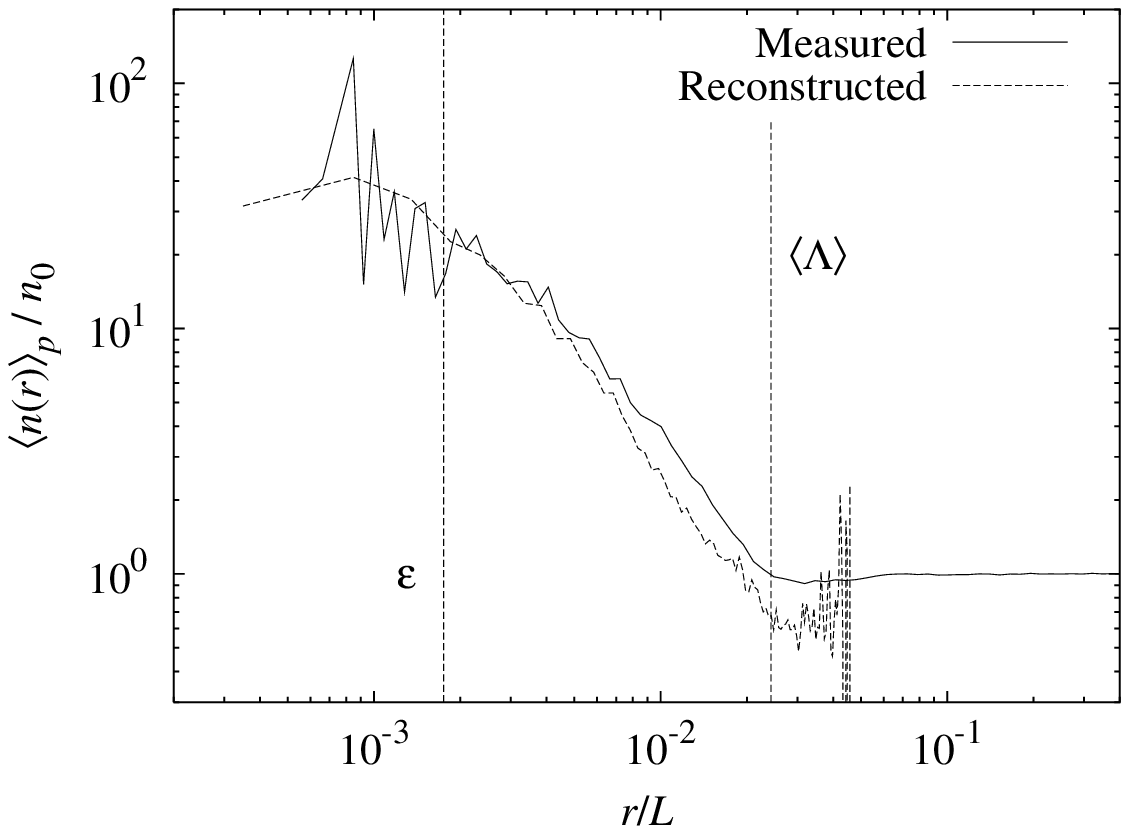}
\caption{Reconstruction of 
the conditional density from $\omega(r)$ at $t=3\,\tau$ in the shuffled lattice.
\label{omega_gamma_sl_30}}
\end{figure}

\begin{figure}
\includegraphics[width=9cm,angle=0]{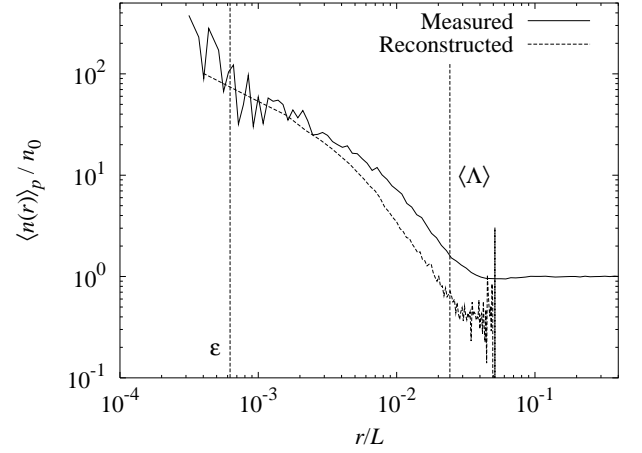}
\caption{Reconstruction of the 
conditional density from $\omega(r)$ at $t=4\,\tau$ in the shuffled lattice.
\label{omega_gamma_sl_40}}
\end{figure}

The Figs \ref{omega_gamma_sl_0} to \ref{omega_gamma_sl_40} show the
reconstructed conditional density (by using the NN probability
distribution) and the one measured directly in the simulation. As for
the Poisson case, one sees that the first structures observed via the
conditional density are only due to a change of the NN probability
distribution. Of course the dynamics of a particle with its NN are not
described in the same way as for the Poisson case. The force on a
particle cannot be approximated by the one from its NN but the latter
seems to be sufficiently important to give the direction of the
particle displacement. Another interesting point is the fact that
there are two phases in the clustering.  This can be seen in
Fig.~\ref{gamma_sl}: between $t_0=0$ and $t_1=\tau$ almost nothing
happens while between $t_1=\tau$ and $t_2=2\,\tau$ the correlations
are quickly developed. As $t_2-t_1=\tau$ is the typical time-scale for
two isolated particles, separated by a distance of order $l$, to
fall on each other, this seems to show that this brutal change is a
sign of such a behavior.


In order to verify this argument we have done a simple test: we have
run the simulation again but with a modified integrator which, for a
given particle, calculates the force acting on it only from its $n$
NNs, $n$ being an integer identical for all the particles, which can be
chosen arbitrarily and changed during the simulation. For our study
what we have done exactly is the following:
\begin{enumerate}
\item at $t=0$, the integrator finds the 6 NNs of each particle;
\item it starts to evolve the system up to $t=\tau$ but at each time
  step, the force on a particle is due only to the 6 particles, which
  were its 6 NNs at $t=0$;
\item at $t=\tau$, the  integrator 
  finds the NN neighbor of each particle;
\item it continues the evolution up to $t=2\,\tau$, the force on a
  particle being now only the one from the particle which was its NN 
  at $t=\tau$.
\end{enumerate}

\begin{figure}
\includegraphics[width=9cm,angle=0]{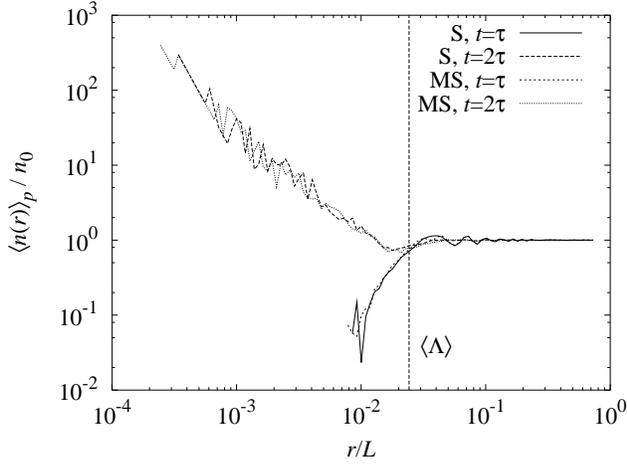}
\caption{Conditional density in the simulation (S) and in the modified
  simulation (MS) for the shuffled lattice.
\label{gamma_6nn}}
\end{figure}

In Fig.~\ref{gamma_6nn} we show the result which confirms our
assumption: between $\tau$ and $2\,\tau$, the dynamics is driven by NN
interaction.  Furthermore one can see that between $0$ and $\tau$,
what matters for a particle is the force from its 6 NNs chosen for the
reason that in a perfect lattice, for a given particle, there is not a
single NN but there are 6 NNs, all at the same distance $\av\Lambda$.
In the case of the lattice, these 6 particles are in a perfect
symmetric configuration around the center particle (this is the case
for all the particles when considered as centers). This implies that
the force resulting from these particles cancels.  In a shuffled
lattice, as long as the parameter $a_s$ is smaller than one, this
remains the case: even if there is a single NN for each particle,
there are always 5 others particles which are almost at the same
distance as the NN.
The simple test presented shows actually that the force from these 6
particles is what matters for the evolution of the correlations in the
system between $t=0$ and $\tau$ and the force from more distant
particles is negligible.

\begin{figure}[!ht]
\includegraphics[width=3cm]{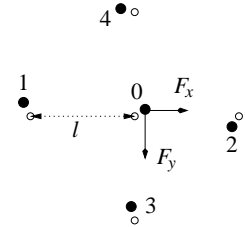}
\caption{Force in a shuffled lattice. The small circles
  ({\small$\circ$}) represent the lattice points while the black dots
  ($\bullet$) represent the particles.  \label{fig:shuffled_lattice}}
\end{figure}

Some simple calculations show actually that the force on a particle in
a shuffled lattice is approximatively given by
\begin{equation}
F_{\text{s}} =  2 \sqrt{3}\; \frac{a_s}{l^2}\, , \label{eq:av_force}
\end{equation}
assuming $Gm^2=1$.  Looking at Fig.~\ref{fig:shuffled_lattice}, one
has for instance that the squared distance $r_{01}$ between the central
particle $0$ and particle $1$ is given by 
\begin{equation} 
r_{01}^2 = l^2\left[
  \left(1-\frac{\varepsilon_{1,x}-\varepsilon_{0,x}}{l}\right)^2 + 
  \sum_{k=y,z}\left(\frac{\varepsilon_{1,k}-\varepsilon_{0,k}}{l}\right)^2 \right]
\end{equation}
where
$\varepsilon_i=(\varepsilon_{i,x},\varepsilon_{i,y},\varepsilon_{i,z})$
is the displacement of the \emph{i}th particle with respect to its
lattice point. Supposing that these displacements are small compared
to $l$, one finds that the force on the central particle from particle
$1$ is
\begin{subequations}
\begin{align}
F_{1,x} & =  \frac{
  2(\tilde{\varepsilon}_{0,x}-\tilde{\varepsilon}_{1,x})-1}{l^2} + \mathcal{O}(\tilde\varepsilon^2)\; , \\
F_{1,k} & = \frac{
  \tilde{\varepsilon}_{1,k}-\tilde{\varepsilon}_{0,k}}{l^2} + \mathcal{O}(\tilde\varepsilon^2) \qquad \text{for
  } k=y,z\; ,
\end{align}
\end{subequations}
with $\tilde{\varepsilon}_{i,k} = \varepsilon_{i,k}/l$.  Making now
the sum over the 6 particles around and averaging on all the
$\varepsilon_{i,k}$ which are random variables going from $-\delta$ to
$\delta$, one obtains $\av{F_x}=\av{\sum_1^6 F_{i,x}}=0$ and a
variance $\av{F_x^2} = 4\delta^2/l^6 $. This gives for the total
squared force
\begin{equation}
\av{F^2}= \av{F_x^2 +F_y^2 +F_z^2}=\frac{12\delta^2}{l^6}
\end{equation}
whose square root is given by Eq.~\eqref{eq:av_force}. The force from
the NN is given approximatively by $F_{\text{NN}}\approx l^{-2}$ which
shows that the real force is roughly
\begin{equation}
F_{\text{s}}\approx 2\sqrt{3}\; a_s F_{\text{NN}}.
\end{equation}
One can estimate a time scale $t_\text{s}$ defined by the relation
$l/2=Gm F _{\text{s}} t_\text{s}^2 /2 $~ which is an approximative
upper bound for the time scale needed by two NN particles to fall on
each other:
\begin{equation}
t_\text{s}=\sqrt{\frac{2\pi}{\sqrt{3} \; a_\text{s} }}\;
\frac{1}{\sqrt{4\pi G \rho_0}} \approx 1.7 \, \frac{\tau}{
\sqrt{a_\text{s}}}
\end{equation}
with $\tau$ given by Eq.~\eqref{tau}. Some simple numerical tests
performed by varying $a_s$ show that the real time scale is closer
to
\begin{equation}
\tau_\text{s} \approx \, \frac{\tau}{ \sqrt{a_\text{s}}} \lesssim t_s
\;.
\end{equation}
In our case, with $a_\text{s}=0.25$, this gives $\tau_\text{s} \approx
2 \,\tau$ which is indeed in good agreement with the simulation.

In summary, while in a Poisson simulation, the correlations are made
directly from the interaction between NN, in a shuffled lattice, there
is a first phase in which a particle interacts mainly with his 6
NNs. This phase is characterized by strong anti-correlations which are
slowly destroyed. This is then followed by a second phase in which some
positive correlations are rapidly developed under some dynamics driven
by NN interactions.


\subsection{CDM simulation}

Finally, we study a CDM cosmological simulation which has been done by
the Virgo Consortium~\cite{virgo}.  This simulation is representative
of many other cosmological simulations as their parameters, their
initial particle configurations and their small scales properties are
more or less always the same. The following discussion should
therefore apply to other cosmological simulations of CDM type.  
Compared to the simulations we have done, this cosmological simulation is
different on two points. Firstly there is space expansion. Secondly the
initial conditions (IC) are very elaborated. 
This last point needs some
explanations. 

The goal of this simulation is to study the evolution of
a gravitating fluid made of CDM particles with particular initial
correlations.  As already mentioned, the particles in the simulation
do not represent CDM particles but are kind of clouds of CDM whose
mass are of the order of a galaxy. This discretization of the fluid
introduces some effects which are reduced by putting initially the
particles in a particular way. The trick is to create first a
distribution where the force on a particle is almost zero.  In the
Virgo case, this is done by running the integrator used for the
simulation on a Poisson distribution with a negative gravity constant
during a while. The distribution obtained is characterized by the fact
that the main part of the force on a particle comes from large scales
mass fluctuations. The contribution from nearby particles is
negligible.  Note that the use of a repulsive gravity gives a
behavior similar to a \emph{one component plasma}~\cite{lebo}.

On this new distribution, it is necessary to apply a correlated
displacement which would transform a continuous and perfectly uniform
distribution into the expected CDM fluid, i.e.  with a power spectrum
on relatively large scales~\footnote{It is not the aim of this
simulation to consider the small $k$ region where $P(k)\sim k$.} behaving as $P(k)
\sim k^n$ with $-3< n < -1$. This displacement field is applied by
using the Zeldovich approximation which also fixes the initial
velocity of each particle as a function of its displacement. The
distribution obtained is therefore correlated at all scales and has
some small initial velocity~\footnote{Note that the velocities are
small~\cite{bsl02}.  This is why afterwards we dare to compare this
simulation with our initially static simulations.}.

Note that as the pre-initial distribution has super-homogeneous
properties (as a lattice or a one-component plasma \cite{GJS,lebo})
there are two main points to be considered:
(\emph{i}) on small scales the distribution continues to have 
the excluded volume feature typical of super-homogeneous
systems,
(\emph{ii}) on large scales the correlations properties
are given by a complex combination of the 
pre-initial correlations (which are long-ranged) and by the 
correlations imposed by the displacement field. Whether 
the resulting fluctuations field has the same small-scales 
properties of the 
CDM continuous distribution is questionable 
\cite{bsl02,bsl03,dk03,andrea}. 
However here we are interested only on the small-scales
features which have the clear imprint of the pre-initial
super-homogeneous distribution very similar, as we discuss below,
to the shuffled lattice.

This simulation is made with $N=256^3$ particles in a box of size
$L=239.5 \text{ Mpc}/h$ (where $0.5\lesssim h \lesssim 1$ is the
dimensionless Hubble constant). The masses are such that $\Omega=1$
and it should represent a standard CDM model. The softening is
$\epsilon = 0.036 \text{ Mpc}/h$ which gives $\eta \approx 25$. This
simulation goes from a redshift $z=50$ to $z=0$.

\begin{figure}
  \includegraphics[width=9cm,angle=0]{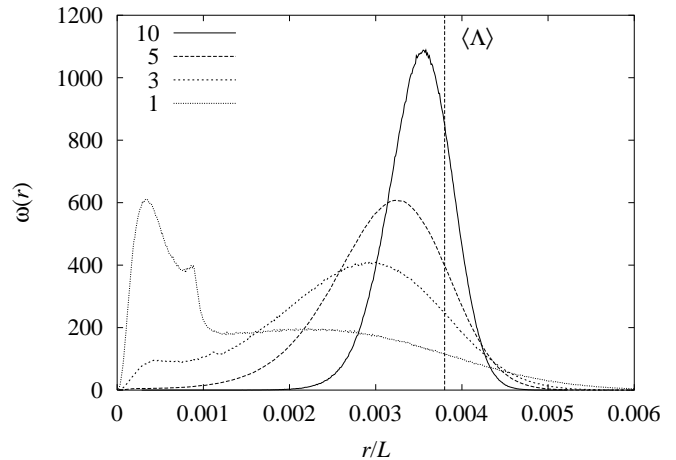}
  \caption{Evolution of the NN probability distribution
    in the CDM simulation.
    The times are given by the redshift $z$. \label{fig:virgo_omega} }
\end{figure}

\begin{figure}
  \includegraphics[width=9cm,angle=0]{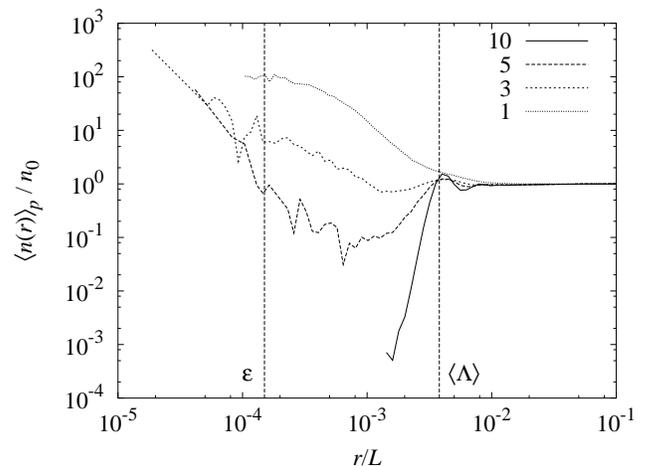}
  \caption{Evolution of the conditional density
    in the CDM simulation.
    The times are given by the redshift $z$. \label{fig:virgo_gamma} }
\end{figure}

\begin{figure}
  \includegraphics[width=9cm,angle=0]{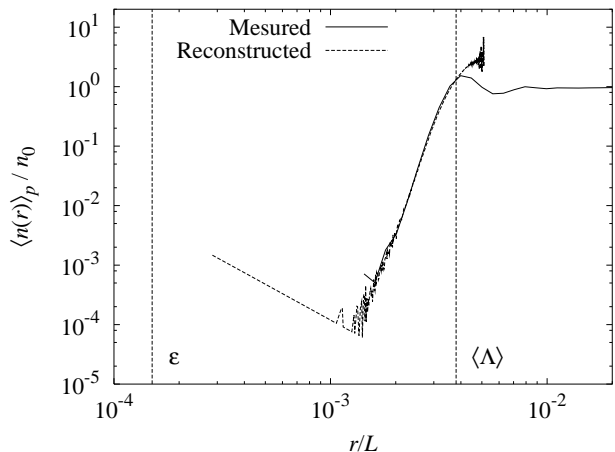}
  \caption{Reconstruction of the conditional 
density from $\omega(r)$ in the CDM
  simulation  at $z=10$.
    \label{omega_gamma_virgo_z10}}
\end{figure}

\begin{figure}
  \includegraphics[width=9cm,angle=0]{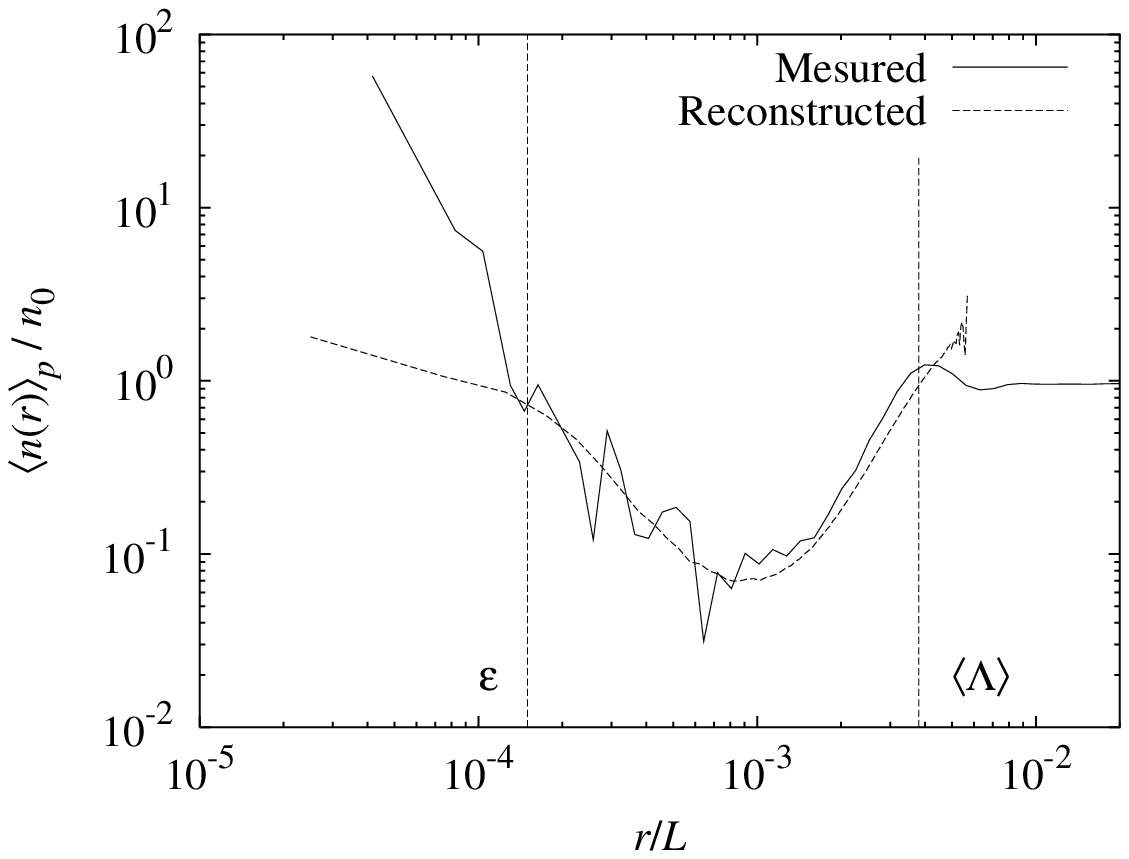}
  \caption{Reconstruction of the 
conditional density from $\omega(r)$ in the CDM
  simulation  at $z=5$. Note that the discrepancy at scales below
  $\epsilon$ comes from a too small statistics on the measured
  conditional density.
    \label{omega_gamma_virgo_z5}}
\end{figure}

\begin{figure}
  \includegraphics[width=9cm,angle=0]{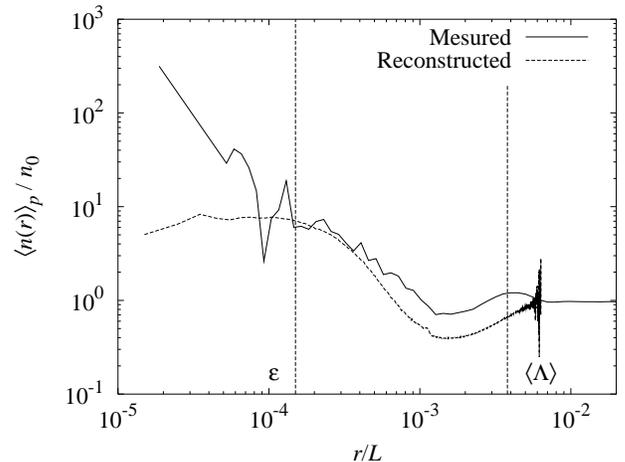}
  \caption{
    Reconstruction of the conditional 
density from $\omega(r)$ in the CDM
  simulation  at $z=3$. Note that the discrepancy at scales below
  $\epsilon$ comes from a too small statistics on the measured
  conditional density.
    \label{omega_gamma_virgo_z3}}
\end{figure}

We have measured the conditional density and the NN probability
distribution. With the latter we have used the approximation based on
NN probability distribution to compute the conditional density.  The
results are shown in
Figs~\ref{omega_gamma_virgo_z10}-\ref{omega_gamma_virgo_z3} while the
evolution of the NN probability distribution and the conditional
density are shown in Figs~\ref{fig:virgo_omega} and
\ref{fig:virgo_gamma}.

The first striking feature that we note is that the evolution is very
similar to the shuffled lattice case.  The conditional density from
being anti-correlated distribution develops positive power-law like 
correlations at scales smaller than $\LL$.
  
This evolution of the correlations is well described  by using
the NN probability distribution, which means that these correlations
are simply due to correlations between NNs. In \cite{bjsl02}, we have
already analyzed this simulation. We had observed that correlations
started at the smallest scales of the system, i.e. $\epsilon <
r < \LL$. Now with the relation between the NN probability
distribution and the conditional density, we can make this observation
more accurate: the ``correlations at the smallest scales'' are
actually correlations between NNs.  As in \cite{bjsl02}, we can again
raise the question of whether these correlations are due to some
interactions between NNs or are a ``large scale'' effect, i.e. a
consequence of the initial velocity field and the acceleration of the
particles under the gravity of large scales mass fluctuations. This
large scale effect would be what we expect from fluid dynamics.

The main point in \cite{bjsl02} and \cite{slbj03} was the kind of
universality of the correlations developed in different gravitating
systems of particles, among them this CDM simulation, a Poisson and
a shuffled lattice. Now we can add that the first correlations are
exactly of the same kind in all these simulations, namely NN
correlations. As in a Poisson and a shuffled lattice, the
discretization plays an important role in the creation of these
correlations, this would suggest that it is the case for the CDM
simulation.

\section{Conclusions}

The fundamental relation used in this paper is Eq.~\eqref{omega1}. It
relates the NN probability distribution $\omega(r)$ to the conditional
density $\av{n(\ve{r})}_p$ at scales of the order of the average
distance between NNs as long as most of the particles have a clear NN.
By checking if this relation holds in a simulation, we get an
interesting information on the nature of the correlations: are they
only due to NN correlations or do they show the existence of
structures made of many particles.

In three simulations that we have considered, Poisson, shuffled
lattice and CDM, which are high-resolution ones ($\eta \gg 1$), we
have seen that this relation holds at early times showing that the
correlations grow by being initially only NN correlations.  In another
simulation, Poisson with large softening such that $\eta \ll 1$, we
have seen that this is not the case anymore.  In this simulation, the
first correlations are due directly to the formations of large
structures --- i.e.  larger than the typical distance between
NNs --- containing more than two particles.

The results for the high-resolution Poisson simulation and for the
shuffled lattice has encouraged us to push the analysis a bit further.
In the Poisson case, using the relation~\eqref{omega1} and considering
the following facts:
(\emph{i})  the force on a particle is mainly due to its NN,
(\emph{ii}) for more than half of the particles, two particles are mutually
  NN;
we could treat the system as a set of isolated two-body
systems. Knowing the initial NN probability distribution and using the
Liouville theorem, it has been then possible to find the early
evolution of the correlations with quite a lot of precision.

In the shuffled lattice, we have observed that at the beginning the
situation was more complex than in the Poisson case. Due to the
approximative symmetries, the early evolution involves interactions
between more than two particles.  Actually, in this case, instead
of having a single NN for each particle, there are six particles which
lie at almost the same distance: such a situation changes the small
scale behavior of the force on an average particle by introducing a
compensation, which is exact and gives a null force only when the
lattice case is considered.  However, the result is similar to the
Poisson case: the formation of correlations between NNs. At a later
time, the situation becomes exactly identical to the Poisson case as
the system behaves like a set of isolated two-body systems. The
consequence is a rapid growth of the correlations between NNs.

For the CDM simulation, we have not tested whether the evolution could be
explained at a certain time by NN interactions. This is a really
important question because this simulation is supposed to describe a
fluid. The particles are not meant to describe particles but mass
tracers: they should follow the flow due the gravity from the large
scales mass fluctuations. If this simulation would have the same
dynamics as the Poisson and shuffled lattice, that is that it could be
explained by NN interactions during a small amount of time, this would
clearly show that the fluid is not well simulated as the evolution
would be influenced by the discrete nature of these particles
resulting from the discretization of the fluid and which 
would therefore not exist in a real fluid.  This would then requires some
careful studies in order to understand how these effects influence the
later evolution.

In some previous papers~\cite{bjsl02,slbj03}, we had already raised
these questions, after having observed the kind of universal
correlations developed in different simulations all characterized by
their particle based dynamics. In some recent
papers~\cite{moore03,bk03}, some others authors have also analyzed the
influence that these particles could have on the evolution but on the
consequences of close encounters between these particles. Their
conclusions were that it has an influence on the density profile of
the clusters.

With this paper we have tried to bring a new element to the
understanding of what happen in such several high resolution
simulations, including the cosmological CDM one,  by showing the
nature of the first correlations developed but we also raise some new
questions which should clearly deserve further studies.  From our
results we now argue for three conclusions about the nature of
clustering in the non-linear regime observed in these NBS.  With
respect to cosmological NBS, we conclude that the exponent
characterizing the non-linear clustering observed has essentially
nothing to do with (\emph{i}) the expansion of the Universe, or (\emph{ii}) the
nature of the small initial fluctuations imposed in the IC.  We
further present evidence for the qualitative description of the
dynamics driving this clustering given in \cite{bottaccio} based on
the Poisson case, and in \cite{bjsl02} based on a similar analysis of
the CDM simulation: (\emph{iii}) the non-linear clustering develops from
the large fluctuations intrinsic to the particle distribution at small
scales (specifically around the smallest resolved scale
$\epsilon$). In particular we show here that the exponent
characterizing it can be seen to emerge at early times in the
simulations when the evolution is well approximated as being due only
to the interactions between NN particles.

A more quantitative description of this dynamics is
evidently needed, with the principal goal of understanding 
the specific value observed of the exponent. In the cosmological 
literature (see e.g. \cite{peebles}) 
the idea is widely dispersed that 
the exponents in non-linear clustering are related to that of 
the initial power-spectrum of the small fluctuations in the CDM fluid,
and even that the non-linear two-point correlation can be considered
an analytic function of the initial two-point correlations
\cite{hamilton,pd96} (although, see \cite{saslaw} where more emphasis is 
put on the tendency for IC to be washed out in the 
non-linear regime).  The models used to explain the behavior 
in the non-linear regime usually involve both the expansion 
of the Universe, and a description of the 
clustering in terms of the evolution of a continuous fluid. We 
have argued that the exponent is universal in a very wide 
sense, being common to the non-linear clustering observed 
in the non-expanding case. It would appear that the framework for 
understanding the non-linear clustering must be one in which 
discreteness (and hence intrinsically non-analytical behavior 
of the density field) is central, and that the simple context
of non-expanding models should be sufficient to elucidate
the essential physics. 
Note that we have not discussed here the {\it amplitude} of the 
correlation function, and in particular how it evolves in time,
which is directly related to the time evolution of the scale of 
non-linearity. This is where the fluctuations at large 
scales, which are different in the various IC considered, can 
play a role as envisaged in the cosmological context (through the 
linear amplification of power at large scales). 
We will address this question further, again 
considering non-expanding models, in future work.

\bigskip

 We warmly
thank D. Pfenniger at the Observatory of
Geneva for the use of the  \textsc{Gravitor} cluster
to run numerical simulations.
We thank  M. Joyce, A. Gabrielli, L. Pietronero, A. Melott, 
and R. Durrer 
for very useful discussions and comments. 
FSL acknowledges the support of a Marie Curie Fellowship
HPMF-CT-2001-01443 
and the University of Geneva for the kind hospitality.
\bigskip


\begin{thebibliography}{26}
\expandafter\ifx\csname natexlab\endcsname\relax\def\natexlab#1{#1}\fi
\expandafter\ifx\csname bibnamefont\endcsname\relax
  \def\bibnamefont#1{#1}\fi
\expandafter\ifx\csname bibfnamefont\endcsname\relax
  \def\bibfnamefont#1{#1}\fi
\expandafter\ifx\csname citenamefont\endcsname\relax
  \def\citenamefont#1{#1}\fi
\expandafter\ifx\csname url\endcsname\relax
  \def\url#1{\texttt{#1}}\fi
\expandafter\ifx\csname urlprefix\endcsname\relax\def\urlprefix{URL }\fi
\providecommand{\bibinfo}[2]{#2}
\providecommand{\eprint}[2][]{\url{#2}}

\bibitem[{\citenamefont{Peebles}(1980)}]{peebles}
\bibinfo{author}{\bibfnamefont{P.~J.~E.} \bibnamefont{Peebles}},
  \emph{\bibinfo{title}{The Large-Scale Structure of the Universe}}
  (\bibinfo{publisher}{Princeton University Press}, \bibinfo{year}{1980}).

\bibitem[{\citenamefont{Saslaw}(2000)}]{saslaw}
\bibinfo{author}{\bibfnamefont{W.~C.} \bibnamefont{Saslaw}},
  \emph{\bibinfo{title}{The Distribution of the Galaxies}}
  (\bibinfo{publisher}{Cambridge University Press}, \bibinfo{year}{2000}).

\bibitem[{\citenamefont{Melott}(1990)}]{melott90}
\bibinfo{author}{\bibfnamefont{A.}~\bibnamefont{Melott}},
  \bibinfo{journal}{Comments Astrophys.} \textbf{\bibinfo{volume}{15}},
  \bibinfo{pages}{1} (\bibinfo{year}{1990}).

\bibitem[{\citenamefont{Kuhlman et~al.}(1996)\citenamefont{Kuhlman, Melott, and
  Shandarin}}]{kuhlman96}
\bibinfo{author}{\bibfnamefont{B.}~\bibnamefont{Kuhlman}},
  \bibinfo{author}{\bibfnamefont{A.}~\bibnamefont{Melott}}, \bibnamefont{and}
  \bibinfo{author}{\bibfnamefont{S.~F.} \bibnamefont{Shandarin}},
  \bibinfo{journal}{Astrophys. J.} \textbf{\bibinfo{volume}{470}},
  \bibinfo{pages}{L41} (\bibinfo{year}{1996}).

\bibitem[{\citenamefont{Splinter et~al.}(1998)\citenamefont{Splinter, Melott,
  Shandarin, and Suto}}]{melott}
\bibinfo{author}{\bibfnamefont{R.~J.} \bibnamefont{Splinter}},
  \bibinfo{author}{\bibfnamefont{A.~L.} \bibnamefont{Melott}},
  \bibinfo{author}{\bibfnamefont{S.}~\bibnamefont{Shandarin}},
  \bibnamefont{and} \bibinfo{author}{\bibfnamefont{Y.}~\bibnamefont{Suto}},
  \bibinfo{journal}{Astrophys. J.} \textbf{\bibinfo{volume}{497}},
  \bibinfo{pages}{38} (\bibinfo{year}{1998}).

\bibitem[{\citenamefont{Baertschiger and Sylos~Labini}(2002)}]{bsl02}
\bibinfo{author}{\bibfnamefont{T.}~\bibnamefont{Baertschiger}}
  \bibnamefont{and}
  \bibinfo{author}{\bibfnamefont{F.}~\bibnamefont{Sylos~Labini}},
  \bibinfo{journal}{Europhys. Lett.} \textbf{\bibinfo{volume}{57}},
  \bibinfo{pages}{322} (\bibinfo{year}{2002}).

\bibitem[{\citenamefont{Baertschiger and Sylos~Labini}(2003)}]{bsl03}
\bibinfo{author}{\bibfnamefont{T.}~\bibnamefont{Baertschiger}}
  \bibnamefont{and}
  \bibinfo{author}{\bibfnamefont{F.}~\bibnamefont{Sylos~Labini}},
  \bibinfo{journal}{Europhys. Lett.} \textbf{\bibinfo{volume}{63}},
  \bibinfo{pages}{633} (\bibinfo{year}{2003}).

\bibitem[{\citenamefont{Dominguez and Knebe}(2003)}]{dk03}
\bibinfo{author}{\bibfnamefont{A.}~\bibnamefont{Dominguez}} \bibnamefont{and}
  \bibinfo{author}{\bibfnamefont{A.}~\bibnamefont{Knebe}},
  \bibinfo{journal}{Europhys. Lett.} \textbf{\bibinfo{volume}{63}},
  \bibinfo{pages}{631} (\bibinfo{year}{2003}).

\bibitem[{\citenamefont{Gabrielli et~al.}(2003)\citenamefont{Gabrielli,
  Jancovici, Joyce, Lebowitz, Pietronero, and Sylos~Labini}}]{lebo}
\bibinfo{author}{\bibfnamefont{A.}~\bibnamefont{Gabrielli}},
  \bibinfo{author}{\bibfnamefont{B.}~\bibnamefont{Jancovici}},
  \bibinfo{author}{\bibfnamefont{M.}~\bibnamefont{Joyce}},
  \bibinfo{author}{\bibfnamefont{J.}~\bibnamefont{Lebowitz}},
  \bibinfo{author}{\bibfnamefont{L.}~\bibnamefont{Pietronero}},
  \bibnamefont{and}
  \bibinfo{author}{\bibfnamefont{F.}~\bibnamefont{Sylos~Labini}},
  \bibinfo{journal}{Phys. Rev.} \textbf{\bibinfo{volume}{D67}},
  \bibinfo{pages}{043506} (\bibinfo{year}{2003}).

\bibitem[{\citenamefont{Gabrielli}(2004)}]{andrea}
\bibinfo{author}{\bibfnamefont{A.}~\bibnamefont{Gabrielli}}
  (\bibinfo{year}{2004}), \bibinfo{note}{in preparation}.

\bibitem[{\citenamefont{Bottaccio
  et~al.}(2002{\natexlab{a}})\citenamefont{Bottaccio, Capuzzo-Dolcetta,
  Miocchi, Montuori, and Pietronero}}]{bottaccio}
\bibinfo{author}{\bibfnamefont{M.}~\bibnamefont{Bottaccio}},
  \bibinfo{author}{\bibfnamefont{R.}~\bibnamefont{Capuzzo-Dolcetta}},
  \bibinfo{author}{\bibfnamefont{P.}~\bibnamefont{Miocchi}},
  \bibinfo{author}{\bibfnamefont{M.}~\bibnamefont{Montuori}}, \bibnamefont{and}
  \bibinfo{author}{\bibfnamefont{L.}~\bibnamefont{Pietronero}},
  \bibinfo{journal}{Europhys. Lett.} \textbf{\bibinfo{volume}{7}},
  \bibinfo{pages}{315} (\bibinfo{year}{2002}{\natexlab{a}}).

\bibitem[{\citenamefont{Mohayaee and Pietronero}(2003)}]{luciano}
\bibinfo{author}{\bibfnamefont{R.}~\bibnamefont{Mohayaee}} \bibnamefont{and}
  \bibinfo{author}{\bibfnamefont{L.}~\bibnamefont{Pietronero}},
  \bibinfo{journal}{Physica A} \textbf{\bibinfo{volume}{323}},
  \bibinfo{pages}{445} (\bibinfo{year}{2003}).

\bibitem[{\citenamefont{Diemond et~al.}()\citenamefont{Diemond, Moore, Stadel,
  and Kazantzidis}}]{moore03}
\bibinfo{author}{\bibfnamefont{J.}~\bibnamefont{Diemond}},
  \bibinfo{author}{\bibfnamefont{B.}~\bibnamefont{Moore}},
  \bibinfo{author}{\bibfnamefont{J.}~\bibnamefont{Stadel}}, \bibnamefont{and}
  \bibinfo{author}{\bibfnamefont{S.}~\bibnamefont{Kazantzidis}},
  \eprint{astro-ph/0304549}.

\bibitem[{\citenamefont{Binney and Knebe}(2002)}]{bk03}
\bibinfo{author}{\bibfnamefont{J.}~\bibnamefont{Binney}} \bibnamefont{and}
  \bibinfo{author}{\bibfnamefont{A.}~\bibnamefont{Knebe}},
  \bibinfo{journal}{Mont. Not. R. Astron. Soc.} \textbf{\bibinfo{volume}{333}},
  \bibinfo{pages}{378} (\bibinfo{year}{2002}).

\bibitem[{\citenamefont{Baertschiger et~al.}(2002)\citenamefont{Baertschiger,
  Joyce, and Sylos~Labini}}]{bjsl02}
\bibinfo{author}{\bibfnamefont{T.}~\bibnamefont{Baertschiger}},
  \bibinfo{author}{\bibfnamefont{M.}~\bibnamefont{Joyce}}, \bibnamefont{and}
  \bibinfo{author}{\bibfnamefont{F.}~\bibnamefont{Sylos~Labini}},
  \bibinfo{journal}{Astrophys. J. Lett.} \textbf{\bibinfo{volume}{581}},
  \bibinfo{pages}{L63} (\bibinfo{year}{2002}).

\bibitem[{\citenamefont{Sylos~Labini et~al.}()\citenamefont{Sylos~Labini,
  Baertschiger, and Joyce}}]{slbj03}
\bibinfo{author}{\bibfnamefont{F.}~\bibnamefont{Sylos~Labini}},
  \bibinfo{author}{\bibfnamefont{T.}~\bibnamefont{Baertschiger}},
  \bibnamefont{and} \bibinfo{author}{\bibfnamefont{M.}~\bibnamefont{Joyce}},
  \eprint{astro-ph/0207029}.

\bibitem[{\citenamefont{Bottaccio
  et~al.}(2002{\natexlab{b}})\citenamefont{Bottaccio, Pietronero, Amici,
  Miocchi, Capuzzo~Dolcetta, and Montuori}}]{bott01}
\bibinfo{author}{\bibfnamefont{M.}~\bibnamefont{Bottaccio}},
  \bibinfo{author}{\bibfnamefont{L.}~\bibnamefont{Pietronero}},
  \bibinfo{author}{\bibfnamefont{A.}~\bibnamefont{Amici}},
  \bibinfo{author}{\bibfnamefont{P.}~\bibnamefont{Miocchi}},
  \bibinfo{author}{\bibfnamefont{R.}~\bibnamefont{Capuzzo~Dolcetta}},
  \bibnamefont{and} \bibinfo{author}{\bibfnamefont{M.}~\bibnamefont{Montuori}},
  \bibinfo{journal}{Physica A} \textbf{\bibinfo{volume}{305}},
  \bibinfo{pages}{247} (\bibinfo{year}{2002}{\natexlab{b}}).

\bibitem[{\citenamefont{Chandrasekhar}(1943)}]{chandra43}
\bibinfo{author}{\bibfnamefont{S.}~\bibnamefont{Chandrasekhar}},
  \bibinfo{journal}{Rev. Mod. Phys.} \textbf{\bibinfo{volume}{15}},
  \bibinfo{pages}{1} (\bibinfo{year}{1943}).

\bibitem[{\citenamefont{Gabrielli et~al.}(2002)\citenamefont{Gabrielli, Joyce,
  and Sylos~Labini}}]{GJS}
\bibinfo{author}{\bibfnamefont{A.}~\bibnamefont{Gabrielli}},
  \bibinfo{author}{\bibfnamefont{M.}~\bibnamefont{Joyce}}, \bibnamefont{and}
  \bibinfo{author}{\bibfnamefont{F.}~\bibnamefont{Sylos~Labini}},
  \bibinfo{journal}{Phys. Rev. D} \textbf{\bibinfo{volume}{65}},
  \bibinfo{pages}{083523} (\bibinfo{year}{2002}).

\bibitem[{\citenamefont{Dorfman}(2001)}]{dorfman}
\bibinfo{author}{\bibfnamefont{J.~R.} \bibnamefont{Dorfman}},
  \emph{\bibinfo{title}{An Introduction to Chaos in Nonequilibrium Statistical
  Mechanics}} (\bibinfo{publisher}{Cambridge University Press},
  \bibinfo{year}{2001}).

\bibitem[{\citenamefont{Gallavotti}(1983)}]{gallavotti}
\bibinfo{author}{\bibfnamefont{G.}~\bibnamefont{Gallavotti}},
  \emph{\bibinfo{title}{The Elements of Mechanics}}
  (\bibinfo{publisher}{Springer-Verlag}, \bibinfo{year}{1983}).

\bibitem[{\citenamefont{\texttt{http://www.mpa-garching.mpg.de/gadget/}}()}]{g%
adget}
\bibinfo{author}{\bibnamefont{\texttt{http://www.mpa-garching.mpg.de/gadget/}}%
}.

\bibitem[{\citenamefont{Binney and Tremaine}(1987)}]{binney}
\bibinfo{author}{\bibfnamefont{J.}~\bibnamefont{Binney}} \bibnamefont{and}
  \bibinfo{author}{\bibfnamefont{S.}~\bibnamefont{Tremaine}},
  \emph{\bibinfo{title}{Galactic Dynamics}} (\bibinfo{publisher}{Princeton
  University Press}, \bibinfo{year}{1987}).

\bibitem[{\citenamefont{Jenkins et~al.}(1998)\citenamefont{Jenkins, Frenk,
  Pearce, Thomas, Colberg, White, Couchman, Peacock, Efstathiou, and
  Nelson}}]{virgo}
\bibinfo{author}{\bibfnamefont{A.}~\bibnamefont{Jenkins}},
  \bibinfo{author}{\bibfnamefont{C.~S.} \bibnamefont{Frenk}},
  \bibinfo{author}{\bibfnamefont{F.~R.} \bibnamefont{Pearce}},
  \bibinfo{author}{\bibfnamefont{P.~A.} \bibnamefont{Thomas}},
  \bibinfo{author}{\bibfnamefont{J.~M.} \bibnamefont{Colberg}},
  \bibinfo{author}{\bibfnamefont{S.~D.~M.} \bibnamefont{White}},
  \bibinfo{author}{\bibfnamefont{H.~M.~P.} \bibnamefont{Couchman}},
  \bibinfo{author}{\bibfnamefont{J.~A.} \bibnamefont{Peacock}},
  \bibinfo{author}{\bibfnamefont{G.}~\bibnamefont{Efstathiou}},
  \bibnamefont{and} \bibinfo{author}{\bibfnamefont{A.~H.}
  \bibnamefont{Nelson}}, \bibinfo{journal}{Astrophys. J.}
  \textbf{\bibinfo{volume}{499}}, \bibinfo{pages}{20} (\bibinfo{year}{1998}).

\bibitem[{\citenamefont{Hamilton et~al.}(1991)\citenamefont{Hamilton, Kumar,
  Lu, and Matthews}}]{hamilton}
\bibinfo{author}{\bibfnamefont{A.~J.~S.} \bibnamefont{Hamilton}},
  \bibinfo{author}{\bibfnamefont{P.}~\bibnamefont{Kumar}},
  \bibinfo{author}{\bibfnamefont{E.}~\bibnamefont{Lu}}, \bibnamefont{and}
  \bibinfo{author}{\bibfnamefont{A.}~\bibnamefont{Matthews}},
  \bibinfo{journal}{Astrophys. J.} \textbf{\bibinfo{volume}{374}},
  \bibinfo{pages}{L1} (\bibinfo{year}{1991}).

\bibitem[{\citenamefont{Peacock and Dodds}(1996)}]{pd96}
\bibinfo{author}{\bibfnamefont{J.}~\bibnamefont{Peacock}} \bibnamefont{and}
  \bibinfo{author}{\bibfnamefont{S.}~\bibnamefont{Dodds}},
  \bibinfo{journal}{Mon. Not. R. Astron.} \textbf{\bibinfo{volume}{280}},
  \bibinfo{pages}{L19} (\bibinfo{year}{1996}).

\end{thebibliography}

\end{document}